\documentclass[a4paper,fleqn,usenatbib]{mnras}

\usepackage{newtxtext,newtxmath}

\usepackage[T1]{fontenc}
\usepackage{ae,aecompl}

\usepackage{graphicx}	\usepackage{amsmath}	\usepackage{amssymb}	\usepackage{pdfpages}
\usepackage{gensymb}
\usepackage{tabularx}
\usepackage{subfig}
\usepackage[figuresleft]{rotating}
\usepackage{wrapfig}
\usepackage[outdir=./]{epstopdf}
\usepackage[nice]{nicefrac}
\usepackage{pdflscape}
\usepackage{cleveref}
\usepackage{float}

\newcommand{\rtt}{$\textrm{R}_{23}$ }

\newcommand{\htwo}{H{\sc ii} }

\newcommand{\stha}{[SII]/H$\alpha$ }

\newcommand{\ntha}{[NII]/H$\alpha$ }

 \usepackage[outdir=./]{epstopdf}
\usepackage[figuresleft]{rotating}

\title[Effects of DIG on Metallicity Gradients]{The Effects of Diffuse Ionized Gas and Spatial Resolution on Metallicity Gradients: TYPHOON Two-Dimensional Spectrophotometry of M83}

\author[Poetrodjojo et al.]{
Henry Poetrodjojo$^{1,2}$,\thanks{E-mail: henry.poetrodjojo@anu.edu.au }
Joshua J. D'Agostino$^{1,2}$,
Brent Groves$^{1,2}$,
Lisa Kewley$^{1,2}$,
\newauthor
\hspace{0.125cm}I-Ting Ho$^{3}$,
Jeff Rich$^{4}$,
Barry F. Madore$^{4}$
and Mark Seibert$^{4}$
\\
$^{1}$Research School of Astronomy and Astrophysics, The Australian National University, Cotter Road, Weston, ACT 2611, Australia\\
$^{2}$ARC Centre of Excellence for All Sky Astrophysics in 3 Dimensions (ASTRO 3D)\\
$^{3}$Max Planck Institute for Astronomy, K{\"o}nigstuhl 17, 69117 Heidelberg, Germany\\
$^{4}$Observatories of the Carnegie Institution of Washington, 813 Santa Barbara St, Pasadena, CA 91101, USA
}

\date{Accepted XXX. Received YYY; in original form ZZZ}

\pubyear{2018}

\begin{document}
\label{firstpage}
\pagerange{\pageref{firstpage}--\pageref{lastpage}}
\maketitle

\begin{abstract}
We present a systematic study of the diffuse ionized gas (DIG) in M83 and its effects on the measurement of metallicity gradients at varying resolution scales. Using spectrophotometric data cubes of M83 obtained at the 2.5m duPont telescope at Las Campanas Observatory as part of the TYPHOON program, we separate the \htwo regions from the DIG using the [SII]/H$\alpha$ ratio, {\sc HIIphot} (\htwo finding algorithm) and the H$\alpha$ surface brightness. We find that the contribution to the overall H$\alpha$ luminosity is approximately equal for the \htwo and DIG regions. The data is then rebinned to simulate low-resolution observations at varying resolution scales from 41 pc up to 1005 pc. Metallicity gradients are measured using five different metallicity diagnostics at each resolution. We find that all metallicity diagnostics used are affected by the inclusion of DIG to varying degrees. We discuss the reasons of why the metallicity gradients are significantly affected by DIG using the \htwo dominance and emission line ratio radial profiles. We find that applying the [SII]/H$\alpha$ cut will provide a closer estimate of the true metallicity gradient up to a resolution of 1005 pc for all metallicity diagnostics used in this study. 
\end{abstract}


\section{Introduction}
The gas-phase metallicity of a galaxy is strongly affected by the processes that occur during the galaxy's evolution. Gas inflows, galaxy mergers and galactic winds are a few examples of events that alter the spatial metallicity distribution.  Measuring the metallicity of a galaxy therefore leads to strong constraints on its growth and formation. Many studies have shown that isolated spiral galaxies exhibit a characteristic metallicity gradient when normalized by the disc scale length \citep{sanchez12,sanchez14,ho15,sanchez-menguiano16,poetrodjojo18}. This implies that galaxies tend to form along the same evolutionary track if they are relatively unaffected by their environment. Conversely, interacting galaxies consistently show significantly shallower metallicity gradients than their isolated counterparts \citep{kewley10,rich12,sanchez14}. This flattening is caused by a combination of many processes which stem from the gravitational interactions between the galaxy pairs. Dilution of the metal-rich centre caused by inflows of pristine gas from the outskirts of a galaxy is an example of how flattening can occur in interacting galaxy pairs \citep{rupke10,kewley10}.
\par
Advances in integral field unit (IFU) spectroscopy allow astronomers to spatially resolve detailed physical properties of individual galaxies. Using these spatially resolved emission line spectra, we are able to produce the metallicity map of a galaxy to great detail. Metallicity maps allow us to view the variations of metallicity within a galaxy and to constrain the metallicity gradients with greater certainty rather than placing limited apertures throughout a galaxy. Recently there have been several large scale IFU surveys such as the Calar Alto Legacy Integral Field Area survey \citep[CALIFA,][]{sanchez12}, the Sydney-Australian-Astronomical-Observatory Multi-object Integral-Field Spectrograph survey \citep[SAMI,][]{allen15,croom15,bryant15} and the Mapping Nearby Galaxies at APO survey \citep[MaNGA,][]{bundy15}. The SAMI and MaNGA surveys aim to observe $\sim3600$ and $\sim10000$ galaxies respectively, through the use of multiplexing technology which allows them to view multiple galaxies at once. Such studies greatly increases the number of observed galaxies, and allow statistical studies of resolved properties like metallicity gradients to be determined \citep{sanchez12,sanchez14,ho15,sanchez-menguiano16,belfiore17,poetrodjojo18}. 
\par
The disadvantage of these large scale IFU surveys is that, because of the wide field needed to sample multiple galaxies at once, they often have seeing-limited spatial resolutions on the order of $\sim 1-2$ kpc. Typical \htwo regions range from ten to hundreds of parsecs \citep{azimlu11,gutierrez11,whitmore11}, much smaller than the resolution of these large IFU surveys. This means that the vast majority of resolution elements of the SAMI and MaNGA data contain a mixture of emission from \htwo regions and surrounding diffuse ionized gas (DIG). Most strong emission line metallicity diagnostics are generated on the assumption that the emission lines are produced purely from \htwo regions. The contamination by DIG causes systematic variations in metallicity, which can cause the metallicity gradient to be steepened or flattened depending on the galaxy and metallicity diagnostic used \citep{zhang17}. 
\par
The diffuse ionized gas, also known as the Warm Ionized Medium (WIM), has long been a region of interest in nearby galaxies. It was first identified by \citet{reynolds84} in the Milky Way and was named the Reynolds layer. Further studies have uncovered it has a significant contribution to the overall luminosity of a galaxy as well as its prevalence in most star-forming galaxies \citep{walterbos94,ferguson96,hoopes96,greenawalt98}. For a comprehensive review of the DIG, see \citet{haffner09}.
\par
DIG is found within the plane of the galaxy disk as well as above and below it, more specifically referred to as the extraplanar diffuse ionized gas (eDIG) \citep{hoopes99,rossa03}. One of the defining features of the DIG are its enhanced emission line ratios including [SII]/H$\alpha$, [NII]/H$\alpha$ and [OI]/H$\alpha$ relative to \htwo regions \citep{hoopes03,madsen06,voges06}, which shifts the DIG towards the LINER and AGN regions of the BPT diagram \citep{baldwin81}. These variations in the emission line ratios combined with its prominent contribution to the total emission line flux, means that it can significantly alter the emission line products of a galaxy such as metallicity, ionization parameter and star-formation rate calculations. In low-resolution observations such as those at high-redshift, large galaxy surveys and aperture measurements, where the DIG can not be isolated due to spatial resolution limitations, a contribution by the DIG is inevitable.
\par
The source of ionizing photons of the DIG is still a debated topic with two likely explanations: leakage of ionizing photons from \htwo regions and ionization by low-mass evolved stars. The spatial correlation of \htwo regions and the DIG suggests leaky \htwo regions as a strong candidate for the source of ionizing photons. Using the Survey for Ionization in Neutral Gas Galaxies (SINGG) sample, \citet{oey07} found an anti-correlation between the fraction of H$\alpha$ surface brightness from the DIG and the overall H$\alpha$ surface brightness. A mean fraction of $0.59\pm0.19$ was found with starburst galaxies ($\Sigma(H\alpha) > 2.5 \times 10^{39}$ erg s$^{-1}$ kpc$^{-2}$) containing the lowest fraction of DIG. However, leaky \htwo regions are unable to fully reproduce the emission line spectrum that we see in the DIG. In particular, \htwo region photon leakage enhances the [SII]/H$\alpha$, [NII]/H$\alpha$ and [OI]/H$\alpha$ emission line ratios, but is unable to produce the [OIII]/H$\beta$ emission line enhancement \citep{zhang17}. It is likely that the DIG is ionized by some combination of ionizing photons produced by leaky \htwo regions and low-mass evolved stars.
\par
Gradient smoothing is another disadvantage caused by the kiloparsec resolution scales of these low-spatial resolution IFU surveys. With large resolution scales of $1-2$ kpc, regions of high metallicity are mixed with regions of lower metallicity, causing the overall smoothing of the metallicity gradient. \citet{yuan13} demonstrated this flattening through annular binning and discussed the implications for measuring metallicity gradients at high redshift using the [NII]/H$\alpha$ metallicity diagnostic \citep{pettini04}. \citet{mast14} used galaxies from the PPAK IFS Nearby Galaxies Survey (PINGS) and degraded the data to different spatial resolutions and showed the flattening of the metallicity gradient at coarser resolution scales, simulating the effects of observing at higher redshifts.
\par
One way to obtain high resolution observations is to observe large nearby galaxies. The large angular size of nearby galaxies allows for an intrinsically higher seeing limited physical resolution. However, because these galaxies tend to occupy a large area of the sky, the typical IFU field of view is far too small to observe the entire galaxy.
\par
TYPHOON/PrISM is a wide field spectrograph survey which aims to produce highly spatially resolved spectrophotometric data of nearby galaxies. Instead of using fibre bundles like large IFU surveys, TYPHOON uses a very long 18$\arcmin$ slit with a width of 1.65$\arcsec$ and steps across the face of the galaxy. By choosing nearby galaxies (z $\leq0.005$), TYPHOON is able to achieve seeing-limited resolutions of up to 2 pc, with a median of 48 pc across their galaxy sample. At these resolution scales we are able to resolve individual \htwo regions without any DIG contamination. When calculating metallicity gradients, we also avoid the smoothing that occurs at coarser resolution scales.
\par
In this paper, we use TYPHOON data to determine the true metallicity gradient of M83/NGC5236 unaffected by DIG contamination or spatial smoothing. We then degrade the data to coarser resolution scales to show the systematic flattening of the metallicity gradient and the implications this will have on large scale IFU surveys. We discuss the effectiveness of applying DIG corrections at low resolution scales when measuring metallicity gradients.
\par
We structure this paper as follows: in Section 2 we summarise the properties of M83 and describe the TYPHOON observations. We describe our procedures for rebinning the native resolution data cube to coarser resolutions and discuss the various metallicity diagnostics we use in Section 3. The results of our study are presented in Section 4 and we discuss the implications our results will have on the interpretation of coarse resolution data products in Section 5. Finally in Section 6 we provide a brief summary and future directions of the research involving the TYPHOON dataset. 
\section{M83}
\subsection{Observations and Properties}
M83/NGC 5236 is a nearby face-on barred spiral galaxy with a galactocentric distance of 4.47 Mpc \citep{tully08} with a redshift of $z=0.001711$. M83 was observed as part of the TYPHOON program using the 2.5m du Pont telescope at the Las Campanas Observatory in Chile. For full information regarding the TYPHOON survey and instrument, see Seibert et al. (in prep). The imaging spectrograph of TYPHOON, Wide Field reimaging CCD (WFCCD), is configured to have a resolving power of approximately R$\approx850$ at $7000\AA$ and R$\approx960$ at $5577\AA$, covering a wavelength range between 3650\AA to $8150\AA$. This allows us to completely separate the [NII] emission lines from H$\alpha$, but does not provide enough resolving power to fit multiple Gaussian components to emission lines. 
\par
A total of 243 observations were spread during 9 nights over 2 observing runs in May 2011 (5 nights) and February 2016 (4 nights). Each slit position was integrated for 600 seconds before being moved by 1.65\arcsec (width of the TYPHOON slit) for the next integration. This process was repeated until the optical disk of M83 was covered, resulting in an image covering an area of $6.7\arcmin\times18\arcmin$. The data is then reduced using standard long-slit data techniques and the final spectrum fit using {\sc LZIFU} \citep{ho16} to produce emission line flux datacubes.
\par
In order to ensure the quality of the data over multiple nights and observing runs, we employ strict observational requirements. We only include data into the final datacube when conditions are photometric with a seeing less than the width of the slit at all times (seeing < $1.65\arcsec$). This means that any emitted light is not being lost due to our narrow slit width. The long length of the slit ($18\arcmin$) means that we are able to utilise the upper and low portions not occupied by the galaxy for calibration purposes. This allows us to calibrate each slit individually, ensuring consistent calibration over the multiple nights which galaxies of the TYPHOON survey are typically observed. 
\par
At the proximity of M83, the $1.65\arcsec$ slit gives us a native resolution of 41 pc. Its proximity and face-on profile have made it one of the most popular and widely studied galaxy to date. This resolution scale allows us to separate \htwo dominated regions from DIG. A table of all intrinsic properties used for this study are given in Table \ref{galinfo}. 
\par
\begin{figure*}
	\begin{tabular}{ccc}
		\centering
	\includegraphics[height=0.4\textheight]{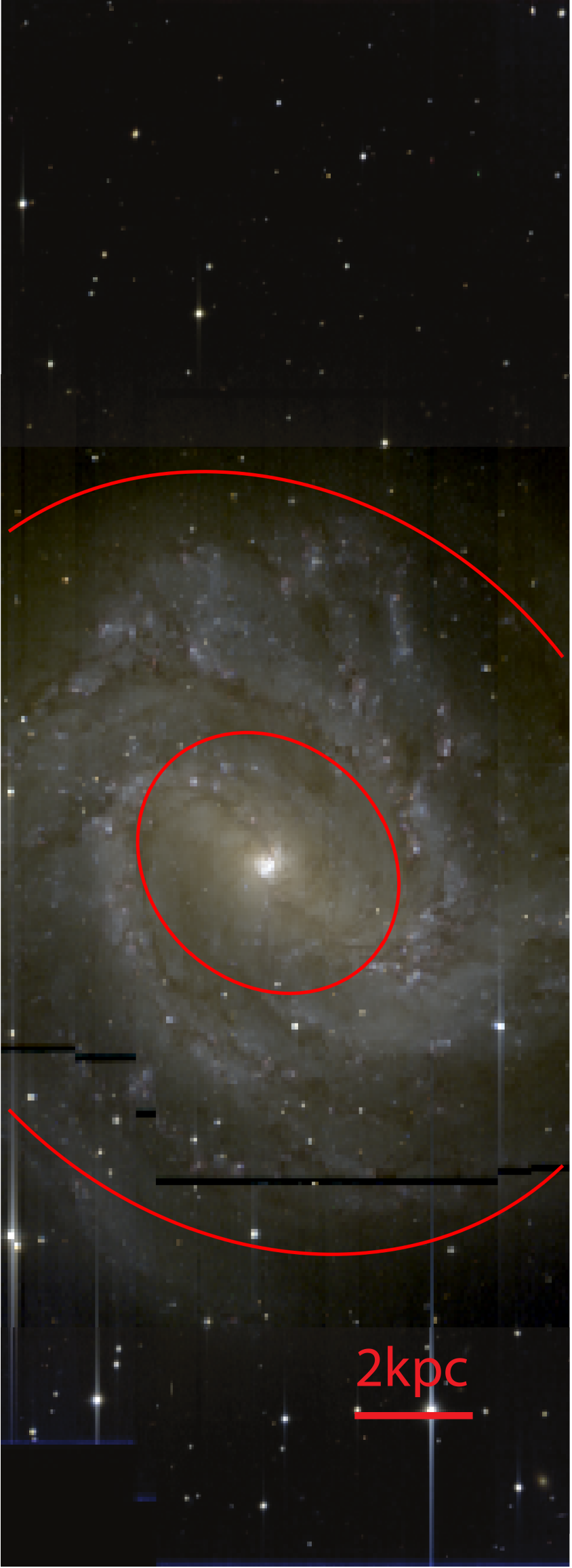}\label{fig:a}&
	\includegraphics[height=0.403\textheight]{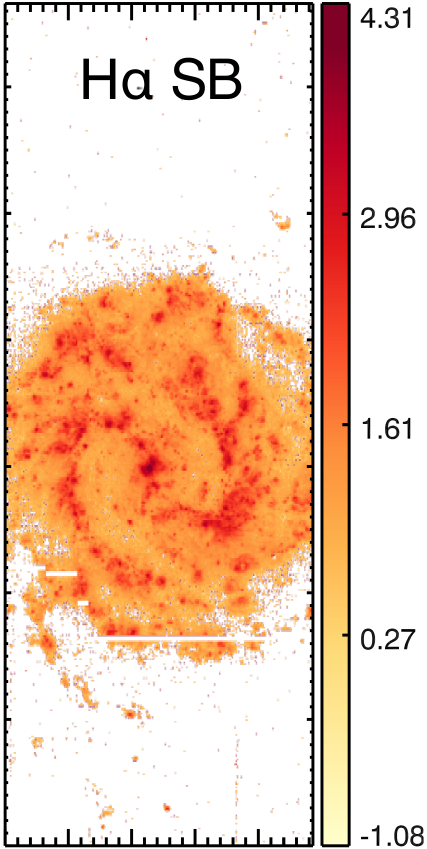}\label{fig:a}&
	\includegraphics[height=0.403\textheight]{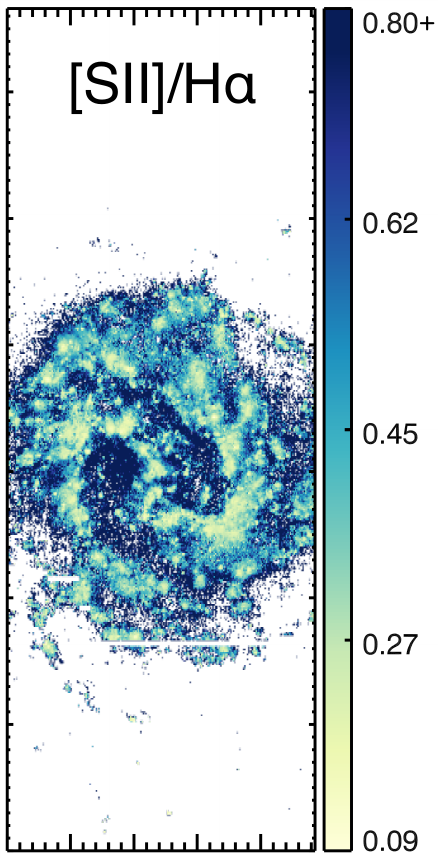}\label{fig:a}\\
	\end{tabular}
	\caption{
\textit{Left}: BVR composite image of M83 created from the TYPHOON datacube. The two red ellipses correspond to 0.5 and 1.5 R$_{e}$ of the galactic disk. \textit{Middle}: H$\alpha$ surface brightness map of M83 extracted by {\sc LZIFU} in units of log(10$^{-17}$ erg/s/cm$^2$/arcsec$^2$). \textit{Right}: The \stha emission line ratio map of M83. Diffuse ionized gas is known to have increased emission line ratios such as \stha and \ntha. As a result, \stha is often used as a tracer for diffuse ionized gas.  
	}
 \label{m83_prop}
\end{figure*}

In Figure \ref{m83_prop} we show the BVR composite image, H$\alpha$ and [SII]/H$\alpha$ line ratio of M83 constructed from the TYPHOON datacube. The bright hot spots that appear along the spiral arms of the H$\alpha$ image indicate the \htwo regions where metallicity diagnostics are valid. We also see significant H$\alpha$ detection in the inter-arm regions corresponding to emission from the DIG. At the native resolution 41pc, we push down to a detection limit of 5.4$\times10^{-17}$ erg/s/cm$^2$/arcsec$^2$ with a mean noise level of 4.2$\times10^{-17}$ erg/s/cm$^2$/arcsec$^2$. As we degrade the spatial resolution, we are able to push towards lower detection limits but the boundary between the \htwo regions and DIG becomes blurred and can no longer be separated.
\par
Observations of the DIG indicate that it is hotter and lower ionization than nearby \htwo regions, with increased [SII]/H$\alpha$ and [NII]/H$\alpha$ ratios. The [SII]/H$\alpha$ ratio is most often used as a DIG indicator as it provides a clean separation between \htwo regions and DIG regions \citep[e.g.][]{blanc09}. The [SII]/H$\alpha$ map in Figure \ref{m83_prop} clearly shows a significant increase in [SII]/H$\alpha$ in the inter-arm regions.

\begin{table}
\centering
\begin{tabular}{c|c|c}
Name & M83/N5236& Reference\\
\hline
Right Ascension & $13^{h}37^{m}0.95^{s}$& \citet{diaz06}\\
Declination & $-29^{\circ}51\arcmin55.5\arcsec$ & \citet{diaz06} \\
Distance & 4.47 Mpc & \citet{tully08}\\
log($M_{*}/M_{\odot}$) & 10.55 & \citet{bresolin16} \\
Effective Radius & 196.7\arcsec & \citet{lauberts89} \\
Position Angle & $44.9^{\circ}$ & \citet{lauberts89} \\
Inclination & $32.5^{\circ}$ & \citet{lauberts89} \\
\end{tabular}
\caption{Fundamental properties of M83 used in this study.}
\label{galinfo}
\end{table}

\subsection{Previous Research}
Due to its proximity, M83 has been a popular subject of many studies over the years. Using strong emission line diagnostics by \citet{kobulnicky99}, a shallow radial metallicity gradient was reported by \citet{bresolin02} by combining their own \htwo region observations and those obtained by \citet{dufour80} and \citet{webster83}. A break in the metallicity gradient is observed in the extended disk of M83 beyond the R$_{25}$ isophotal radius, where the metallicity gradient becomes flat \citep{bresolin09}. Using the "counterpart" method, \citet{pilyugin12} was unable to find solid evidence for a discontinuity between the inner at outer disk of M83. They did however acknowledge that the metallicity gradient did become flatter at the transition point.
\par
With significant deviations in the metallicities measured from different strong emission line diagnostics \citep{kewley08}, \citet{bresolin16} provides stellar metallicity measurements of blue supergiants within the inner disk of M83. They find that the stellar metallicity measurements are in good agreement with the $T_{e}$-based metallicities. With the exception of the ([OIII]/H$\beta$)/([NII]/H$\alpha$) metallicity diagnostic \citep{pettini04}, the strong emission line diagnostics produce significantly different radial profiles to those calculated by the $T_{e}$ method.  
\par
The presence of extra-planar diffuse ionized gas (eDIG) in M83 was detected by \citet{boettcher17} using Markov Chain Monte Carlo methods to decompose the \htwo regions from the eDIG using high spectral resolution observations from the South African Large Telescope (SALT). Due to the relatively low (R$\sim850$) spectral resolution of TYPHOON, we are unable to spectrally decompose the emission line fluxes into multiple components and separate the eDIG from the planar DIG. Although we are unable to separate the eDIG from the DIG that exists in the midplane, \citet{boettcher17} found that for M83, the DIG within the plane of the disk was several orders of magnitudes brighter than the extraplanar component. This means that any \htwo regions that we define in the plane of the disk will be relatively unaffected by the presence of eDIG. 

\section{Method}\label{method}
\subsection{Data Binning}
\begin{figure*}
\begin{centering}
\includegraphics[width=\linewidth]{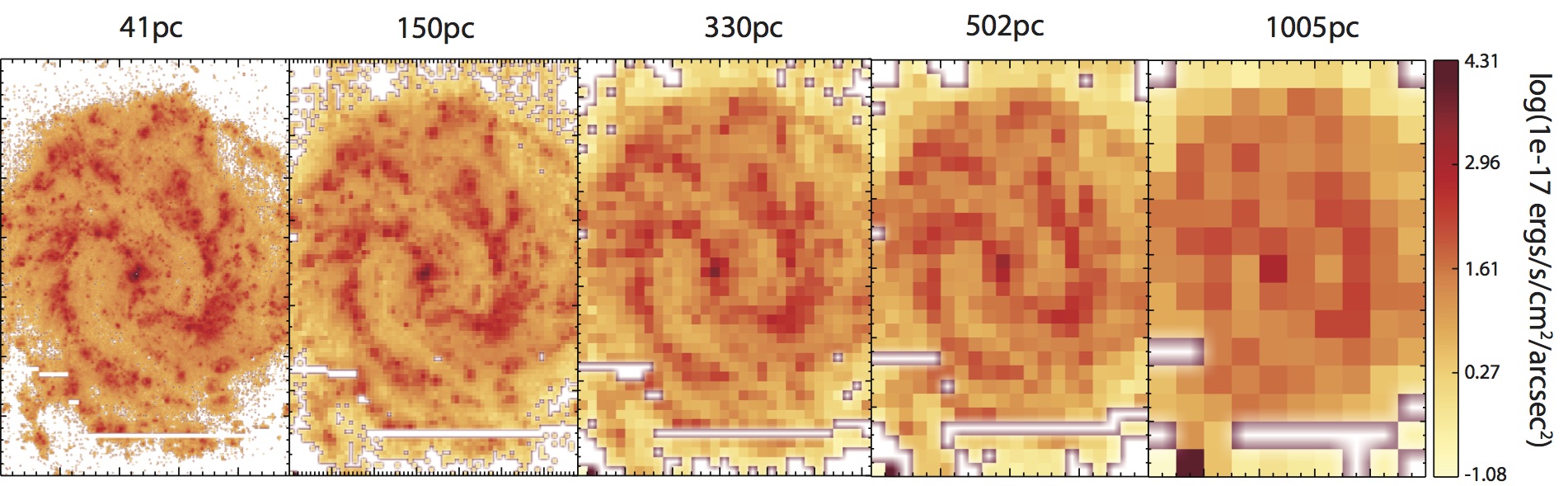}
\end{centering}
\caption{H$\alpha$ map of M83 at different resolution scales. The spatial resolutions from left to right: 41 pc, 150 pc, 330 pc, 502 pc and 1005 pc. The gradual blurring of defined \htwo regions and the merging of the spiral arms with the inter-arm regions can be clearly seen in this figure.}
\label{halpha_res}
\end{figure*}

Figure \ref{halpha_res} shows the H$\alpha$ image at each binned resolution scale. An important feature of Figure \ref{halpha_res} is the gradual blurring of the \htwo regions. At the native resolution of 41 pc, the \htwo regions are clearly distinguishable from the DIG regions as high surface brightness (red) spots mostly distributed along the spiral arms. At the 150 pc resolution, most of the individual \htwo regions are still distinguishable, but some of the more densely packed \htwo regions begin to merge. The resolutions of 330pc and 502 pc are typical of the highest resolution IFU galaxy surveys (eg. CALIFA). At these resolutions, \htwo regions become completely merged with each other and the spiral arm structure becomes the new \htwo region boundary. At 502 pc we also see the boundary between DIG and \htwo regions begin to blur, suggesting increasing DIG contamination. We would expect SAMI and MaNGA on average to achieve resolutions on the scale of our final panel at 1005 pc. Large morphological features such as the strong spiral arms are no longer distinguishable at this level and clearly distinguishing between DIG and \htwo regions is difficult. When working with data at this resolution, DIG contamination is almost certain and thus needs to be taken into consideration. 

\par
Our goal is to find a simple mechanism to remove or minimise the effects of DIG from the determination of metallicity and other emission line derived physical parameters. To do this we must first define the boundary between DIG and \htwo regions. Previous studies have attempted to define DIG by using H$\alpha$ surface brightness \citep{zhang17}, emission line ratios such as [NII]$\lambda6583$/H$\alpha$ and [SII]$\lambda\lambda6717,6731$/H$\alpha$ \citep{blanc09} and advanced \htwo region finding algorithms such as {\sc HIIphot} \citep{thilker00} and {\sc HIIexplorer} \citep{sanchez12}. We apply each of these methods to demonstrate the systematics involved in \htwo and DIG separation on emission line derived physical parameters. We list the number of \htwo and DIG region spaxels for each DIG classification scheme within 1.5 R${_e}$ in Table \ref{htwodignumbers}. Before any of the emission lines are used for defining \htwo regions, we subtract the stellar continuum and fit the strong emission lines using {\sc LZIFU}. We describe the {\sc LZIFU} routine in more detail in Section \ref{LZIFU}. In the following sections we describe our application of each \htwo region classification method.
\subsubsection{Separation by H$\alpha$ Surface Brightness}
Individual \htwo region sizes are typically on the order of tens to hundreds of parsecs \citep{azimlu11,gutierrez11,whitmore11} while the typical resolution of multiplexing IFU surveys tend to have a spatial resolution coarser than $\sim1$ kpc. This means that a clean separation between \htwo regions and DIG emission is difficult. While \htwo regions are typically orders of magnitude brighter than DIG regions, the DIG covers a larger fraction of the galactic disk. However the difference in surface brightness means that a crude separation between \htwo dominated spaxels and DIG dominated spaxels is possible by looking at the surface brightness alone.
\par
\citet{zhang17} used galaxies observed by the MaNGA survey to analyse the effects that the DIG has on emission line ratios in MaNGA observations and its derived properties. With a spatial resolution greater than $\sim1$ kpc, they are unable to cleanly separate the \htwo regions from the DIG. Instead they analyse how the emission line ratios and their products change with H$\alpha$ surface brightness. 
\par
For this paper, we use the method outlined in \citet{kaplan16} to determine the fraction of flux originating from the \htwo regions and the DIG. This method was first developed by \citet{blanc09} and expanded by adding a parameter to allow variation in the DIG surface brightness due to the star-formation distribution. Using the assumption that the brightest spaxels are dominated by \htwo emission and the dimmest spaxels are dominated by DIG emission, \citet{kaplan16} determines a characteristic [SII]/H$\alpha$ emission line ratio for \htwo and DIG regions. The linear distance between the \htwo characteristic [SII]/H$\alpha$ and DIG [SII]/H$\alpha$ ratio determines the percentage of flux originating from each region. For example, if the \htwo  characteristic [SII]/H$\alpha = 0.2$ and the DIG characteristic [SII]/H$\alpha = 0.9$, a spaxel with [SII]/H$\alpha = 0.55$ would have equal contribution of emission line flux from \htwo regions and the DIG.
\par
The fraction (C$_{\textrm{\htwo}}$) of emission originating from the \htwo regions is then mapped to the extinction corrected H$\alpha$ surface brightness (extinction correction method described in Section \ref{extinctioncorrection}) and fit using a function of the form:
\begin{equation}
C_{\textrm{\htwo}}=1.0-\left(\frac{f_0}{f(H\alpha)}\right)^{\beta}
\label{concentration}
\end{equation}
where $f_0$ is the threshold below which a spaxel is completely comprised of DIG and $\beta$ allows for the variation in DIG surface brightness. We fit both parameters using {\sc MPFIT}\citep{markwardt09} and provide all values in Appendix \ref{fitvalues}. We define a \htwo region to be a spaxel in which 90\% of its emission originates from \htwo regions. This corresponds to a H$\alpha$ surface brightness cut-off of $1.86\times10^{-15}$ erg s$^{-1}$ cm$^{-2}$ arcsec$^{-2}$ at the native resolution of 41 pc. 
\subsubsection{[SII]/H$\alpha$ Emission Line Ratio}
\citet{madsen06} found a significant increase in the [SII]/H$\alpha$ and [NII]/H$\alpha$ line ratio in the DIG components of the Milky Way. The \htwo regions had an average value of [SII]/H$\alpha$ = 0.12 and \ntha = 0.23, while the DIG regions had an average value of \stha = 0.38 and \ntha = 0.83. However the \ntha line ratio has a larger scatter than the \stha line ratio, making it less reliable for separating the \htwo and DIG dominated regions.
\par
Using the characteristic [SII]/H$\alpha$ emission line ratio for \htwo and DIG regions determined previously, we linearly map each spaxel to determine the fraction of emission produced by \htwo regions and the DIG. As with the H$\alpha$ surface brightness cut-off, we define a \htwo region to be a spaxel in which 90\% of its emission originates from \htwo regions. A [SII]/H$\alpha$ cut-off of 0.29 is determined at the native resolution of 41 pc. That is, a spaxel with [SII]/H$\alpha < 0.29$ is classified as a \htwo region. Since Equation \ref{concentration} essentially maps the [SII]/H$\alpha$ line ratio to H$\alpha$, classifying a spaxel as a \htwo region or DIG based on the H$\alpha$ surface brightness or [SII]/H$\alpha$ should be exactly the same. However, the differences between the two classification schemes exist because of the scatter around the line of best fit as shown in Figure \ref{41pcs2ha}. 
\subsubsection{\htwo Region Finding Algorithms}
\htwo region finding algorithms provide a way to systematically define the boundaries of \htwo regions, removing the individual biases that may be present when defining by eye. {\sc HIIexplorer} is one such algorithm, used widely in many CALIFA studies \citep{sanchez12,sanchez14,sanchez-menguiano16}. \citet{thilker00} presents {\sc HIIphot}, a robust and systematic method of determining \htwo regions from the H$\alpha$ emission line. {\sc HIIphot} first finds 'seed' \htwo regions and then iteratively grows the seeds until the termination condition is reached. {\sc HIIphot} uses the slope of the H$\alpha$ surface brightness to determine whether it has reached the edge of a \htwo region. The seed threshold and termination conditions of {\sc HIIphot} are user defined, meaning that some subjectivity is inevitably present. We adjust the settings of {\sc HIIphot} to produce \htwo region maps which match our expectations and produce similar regions to the H$\alpha$ surface brightness and [SII]/H$\alpha$ cut-off. The difference between {\sc HIIphot} and the other classification schemes lies within the ability of {\sc HIIphot} to produce smoother boundaries rather than the sharp cut-offs present in the other two classifications.
\begin{table}
	\centering
	\begin{tabular}{c|ccc}
		\multicolumn{3}{c}{Number of \htwo and DIG spaxels} \\
		 Classification Scheme & \htwo & DIG \\
		\hline
		[SII]/H$\alpha$ &     8823 (12\%)&     63288 (88\%) \\
		 {\sc HIIphot} &     4215 (6\%)&     67896 (94\%) \\
		H$\alpha$ SB &     11280 (15\%)&     60831 (85\%) \\

	\end{tabular}
	\caption{
	The number of spaxels considered as \htwo regions or DIG regions within 1.5 R$_{e}$ based on the different \htwo classification schemes.
	}
	\label{htwodignumbers}
\end{table}
\begin{figure*}
\begin{centering}
	\includegraphics[width=\linewidth]{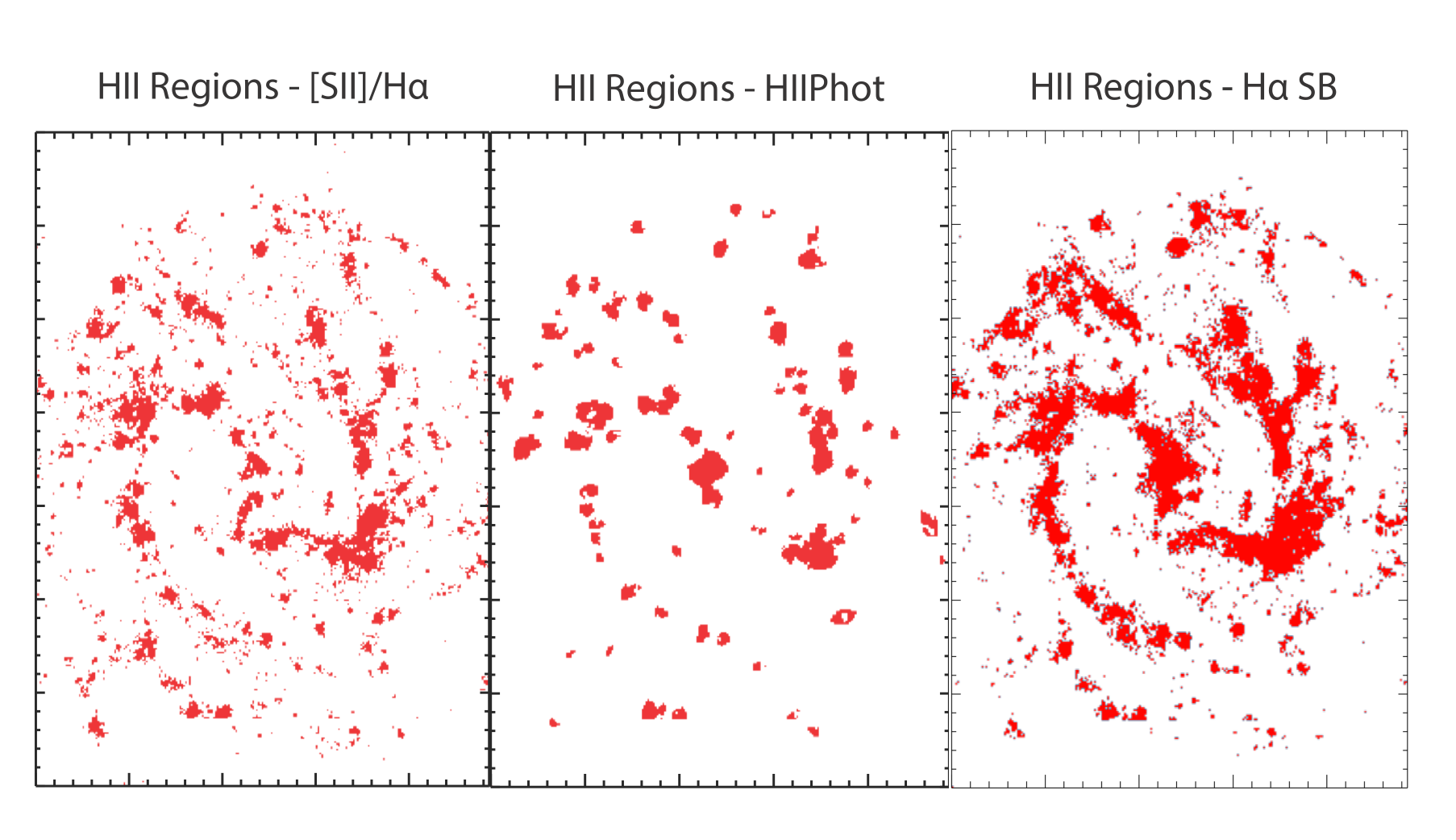}
\end{centering}
	\caption{
		Maps of the three DIG classification schemes used in this paper. \textit{Left}: \htwo region classification using the \stha emission line ratio. [SII]/H$\alpha$ $< 0.29$ is defined as \htwo regions and [SII]/H$\alpha$ $> 0.29$ is defined as DIG. \textit{Middle}: \htwo regions defined using a modified version of {\sc HIIphot} \citep{thilker00}. {\sc HIIphot} determines \htwo regions based on local H$\alpha$ surface brightness profiles. \textit{Right}: \htwo regions determined by a H$\alpha$ surface brightness cut-off. $\Sigma$(H$\alpha$)$ > 1.86\times10^{-15}$ erg s$^{-1}$ cm$^{-2}$ arcsec$^{-2}$ is defined as \htwo regions with anything less being classified as DIG.
	}
	\label{classification}
\end{figure*}
\subsubsection{\htwo Region and DIG Separated Cubes}
\label{LZIFU}
Figure \ref{classification} shows the \htwo region maps for each classification scheme. The [SII]/H$\alpha$ classification scheme produces \htwo regions which appear noisy due to its sharp cut-off. The spiral arm structure is easily extracted using the [SII]/H$\alpha$ emission line ratio and identifies spaxels with low H$\alpha$ surface brightness. The {\sc HIIphot} classification schemes produces much smoother \htwo regions compared to [SII]/H$\alpha$, with \htwo regions appearing rounder with defined borders. A significant difference between regions defined by [SII]/H$\alpha$ and {\sc HIIphot} occurs in the central section of M83. The [SII]/H$\alpha$ classification scheme has a smaller central \htwo region than {\sc HIIphot}, signaling the presence of a high surface brightness region with an enhanced [SII]/H$\alpha$ emission line ratio. Finally, the H$\alpha$ surface brightness cut-off defines the most \htwo region spaxels and bridges all the {\sc HIIphot} \htwo regions together due to its inability to distinguish the presence of high surface brightness DIG regions.
\par
For each \htwo region classification scheme, we create two additional datacubes: one that contains only emission from \htwo regions and one that contains only emission from the rest of the original data cube, which we classify to be DIG. We then rebin the original and each of these new data cubes to lower resolution scales. 
\par
The rebinned \htwo region data cubes are used to simulate how pure spatial smoothing affects observations in the absence of DIG. However the rebinned cubes containing emission outside of the \htwo regions not only contains the DIG emission, but also includes low surface brightness (LSB) regions that fall below the S/N cut at the native resolution. As we rebin to lower resolutions, the LSB regions either merge together with other LSB regions until they have a significant enough S/N or contribute to DIG emission.
\par
Each rebinned data cube is processed by {\sc LZIFU} \citep{ho16}. {\sc LZIFU} extracts total line fluxes for the dominant emission lines by fitting and subtracting the underlying stellar continuum using {\sc ppxf} \citep{cappellari04,cappellari17} and the {\sc miuscat} simple stellar population models \citep{vazdekis12}.  The dominant emission lines are then fit using up to 3 Gaussian profiles with the Levenberg-Marquardt least-squares method \citep{markwardt09}. For this paper, we use the 1-component fits from {\sc LZIFU} for our analysis as including extra Gaussian components does not significantly improve the emission line fits due to relatively low spectral resolution. {\sc LZIFU} returns maps of the flux and flux errors for each emission line, as well as maps of the ionized gas velocity and velocity dispersion and their associated errors (see \citet{ho16} for a detailed explanation of the routine).

\subsection{Metallicity Diagnostics}
\subsubsection{Extinction Correction}
\label{extinctioncorrection}
Before being used in diagnostic ratios, emission lines must be first corrected for attenuation by dust in the interstellar medium (ISM). The attenuation of emission lines is wavelength dependent, meaning that emission line diagnostics that use emission lines with wide wavelength differences are most heavily affected such as N2O2, R$_{23}$ and O32. To correct the emission lines, we apply a S/N cut of 3 and create maps of the observed Balmer ratio, (H$\alpha$/H$\beta)_{\rm obs}$, and solve for E(B-V) by using the relation:
\begin{equation}
E(B-V)=\log_{10}(\frac{(\textrm{H}\alpha/\textrm{H}\beta)_{\rm obs}}{(\textrm{H}\alpha/\textrm{H}\beta)_{\rm int}})/(0.4(k(\rm H\beta)-(k(\rm H\alpha))))
\label{dustext}
\end{equation}
where $(\textrm{H}\alpha/\textrm{H}\beta)_{int}$ is the intrinsic ratio of 2.86 for case B recombination \citep{osterbrock89}. We use the \citet{cardelli89} extinction curve and assume a typical R(V) value of 3.1 to determine $k$ values for H$\alpha$ and H$\beta$. We then use the calculated E(B-V) to determine A($\lambda$) at our emission line wavelengths to de-redden the emission line fluxes. We apply the extinction correction to the N2O2, \rtt and O32 emission line ratios.
\par
We find an average E(B-V) of 0.61 for \htwo regions and 0.80 for DIG regions, corresponding to an A$_{\textrm{v}}$ of 1.89 and 2.48 respectively assuming R(V)=3.1. This agrees with \citet{tomicic17}, who found an increase in the [SII]/H$\alpha$ with A$_{\textrm{v}}$ for resolved spectra in M31. However, a large scatter exists between [SII]/H$\alpha$ and A$_{\textrm{v}}$, leading to the trend being relatively weak. The trend is largely driven by the physical characteristics of the ISM such as gas-phase metallicity and ionization parameter which may vary across a galaxy \citep[e.g.][]{tomicic17}.
\subsubsection{N2O2}
A popular metallicity diagnostic uses the ratio between nitrogen and oxygen emission lines, $\textrm{[NII]}\lambda6583$/$\textrm{[OII]}\lambda3726, \lambda3729$ \citep[N2O2][hereafter KD02]{kewley02}. The main advantage to the N2O2 diagnostic is that because of the similar ionizing potentials of the nitrogen and oxygen species, the diagnostic has very little dependence on the ionization parameter, especially at high metallicity values. Another benefit of the N2O2 diagnostic is that it appears to be one of the metallicity diagnostics least affected by DIG contamination \citep{zhang17}, making it ideal for low resolution data where \htwo regions can not be separated reliably from the DIG. 
\subsubsection{\rtt}
The ([OII]$\lambda\lambda3726,3729$ + [OIII]$\lambda\lambda4959,5007$)/H$\beta$ (\rtt) emission line ratio is one of the most widely used metallicity diagnostics due to its direct use of the oxygen emission lines with a large amount of calibrations using this particular emission line ratio \citep{pagel79,pagel80,edmunds84,mccall85,dopita86,torres-peimbert89,mcgaugh91,zaritsky94,pilyugin00,charlot01,kewley02,kobulnicky04}.
A more complex method of determining the metallicity is by using an iterative method presented in \citet[][hereafter KK04]{kobulnicky04}. The KK04 metallicity diagnostic uses the \rtt line ratio together with the [OIII]$\lambda\lambda4959,5007$/[OII]$\lambda\lambda3726,3729$ O32 emission line ratio to simultaneously determine the metallicity and ionization parameter. The \rtt metallicity diagnostic has a strong dependence on the ionization parameter, making it an ideal diagnostic when determining the metallicity distribution of a galaxy with large ionization parameter variations.
\subsubsection{O3N2}
$(\textrm{[OIII]}\lambda5007$/$\textrm{H}\beta)$/$(\textrm{[NII]}\lambda6583$/$\textrm{H}\alpha)$ \citep[O3N2][hereafter PP04]{pettini04} is another very popular metallicity diagnostic. O3N2 uses the emission lines that are commonly used on the BPT diagram to determine the metallicity. The emission lines involved in the O3N2 metallicity diagnostic are close enough in wavelength that differential extinction (ie. reddening) is minimal. Metallicity varies linearly with the O3N2 emission line ratio, allowing for easy and fast calculations. The only drawback of the O3N2 metallicity diagnostic is that it appears to depend heavily on ionization parameter (Kewley et al. 2019, ARAA, in press), which is not taken into account by PP04. This could lead to potential systematic errors in the metallicity calculations if there is a large variation in ionization parameter throughout the galaxy. 
\subsubsection{N2H$\alpha$}
$\textrm{[NII]}\lambda6583$/$\textrm{H}\alpha$ \citep[N2H$\alpha$][]{storchi-bergmann94,raimann00,denicolo02,pettini04} is another common metallicity diagnostic. Like O3N2, N2H$\alpha$ utilizes emission lines that are close in wavelength to minimise extinction correction, allowing for metallicity measurements where the Balmer ratio can not be reliably determined. Because of the simplicity of only needing two relatively strong emission lines, N2H$\alpha$ is the most popular high redshift metallicity diagnostic. We use the N2H$\alpha$ calibration by PP04 for this paper. However, the N2H$\alpha$ metallicity diagnostic also appears to have a strong dependence on ionization parameter (Kewley et al. 2019, ARAA, in press), resulting in possible systematic errors when using the metallicity diagnostic in its current form. 
\subsubsection{N2S2}
The newest metallicity emission line diagnostic that will be used in our analysis involves the H$\alpha$, [NII]$\lambda6583$ and [SII]$\lambda\lambda6717,6731$ emission lines \citep[N2S2][]{dopita16}. The N2S2 diagnostic uses a combination of the [NII]/H$\alpha$ and [NII]/[SII] emission line ratios in the following sum $\log$([NII]/[SII])$+0.264\times\log($[NII]/H$\alpha)$. N2S2 provides the best of both worlds as the wavelength gap between H$\alpha$, [NII] and [SII] is small enough to ignore for the purposes of extinction correction and is insensitive to changes in the ionization parameter due to the inclusion of the [NII]/[SII] emission line ratio.
\subsection{Error Propagation}
\label{errorprop}
To propagate line flux errors produced by LZIFU through to the metallicity calculations, we simulate 1000 maps for all emission lines used in the calculation. The maps are created such that the fluxes are Gaussian distributed within the LZIFU standard deviation for that emission line. 
\par
Using the simulated line maps, metallicity maps are created for each metallicity diagnostic. The non-linearity of some of the metallicity diagnostics means that the metallicity distributions are not necessarily Gaussian. To represent the spread of metallicity, we determine the distance from the true value to the $16^{\textrm{th}}$ and $84^{\textrm{th}}$ percentiles and calculate the average. This provides us with a measure of the error of the metallicity maps which are then propagated to the gradient errors. 
\section{Results}\label{results}
\begin{figure}
	\includegraphics[width=0.5\textwidth]{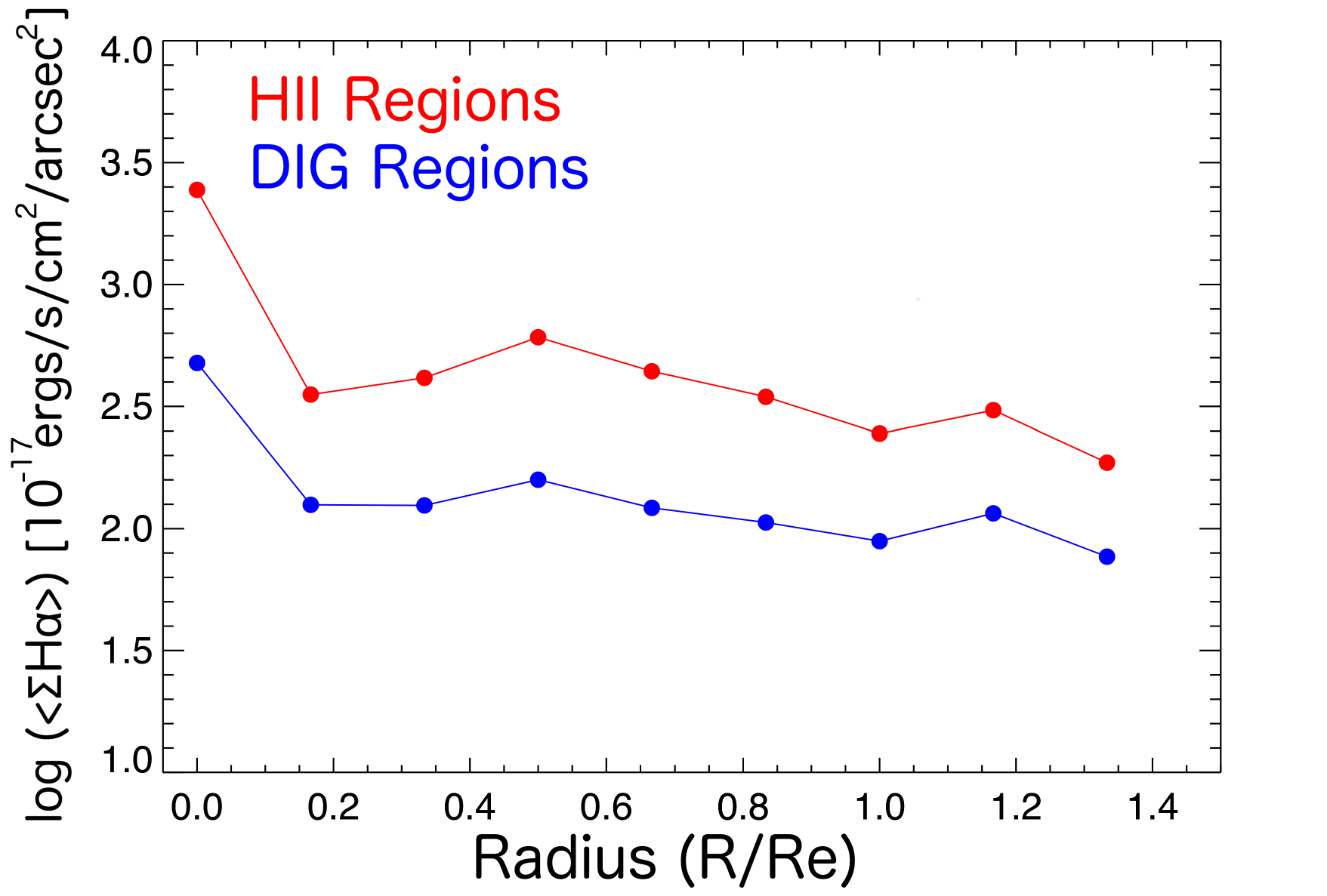}
	\caption{Radial H$\alpha$ surface brightness profiles for \htwo regions and DIG regions using the [SII]/H$\alpha$ \htwo classification scheme. The shape of the radial profiles look almost identical with the \htwo regions being significantly brighter than the DIG regions.}
	\label{HASBProfile}
\end{figure}
\begin{figure}
	\includegraphics[width=0.5\textwidth]{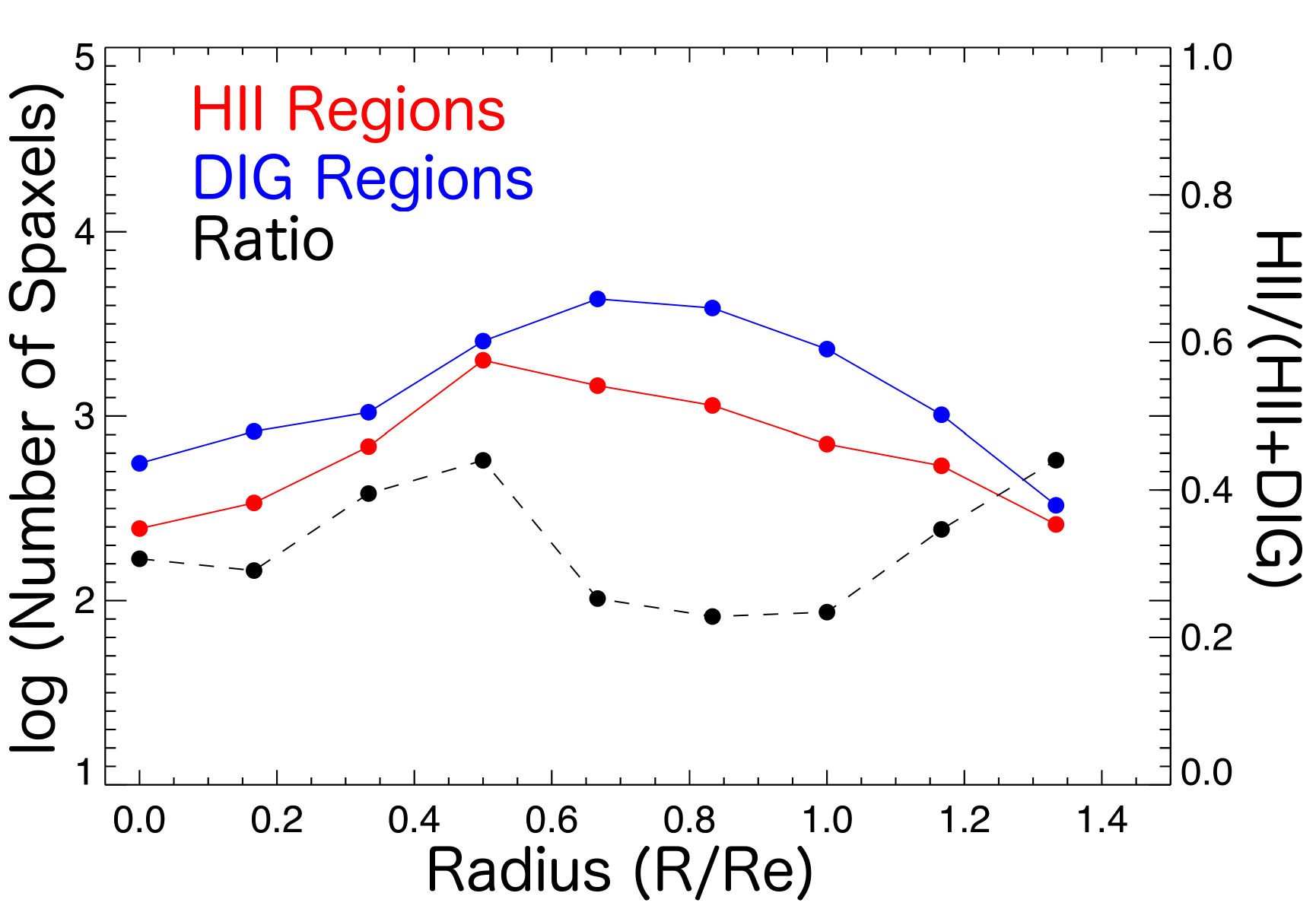}
	\caption{The number of spaxels classified as \htwo and DIG using the [SII]/H$\alpha$ \htwo classification scheme. There are vastly more DIG spaxels than \htwo spaxels at all radii. The number of \htwo region spaxels peaks at R/R$_{e} = 0.6$, corresponding to the location of the spiral arms. The number of DIG spaxels steadily increases at larger radii due to the increase in area but flattens as detection of faint regions becomes difficult.}
	\label{number}
\end{figure}
\begin{figure}
	\includegraphics[width=0.47\textwidth]{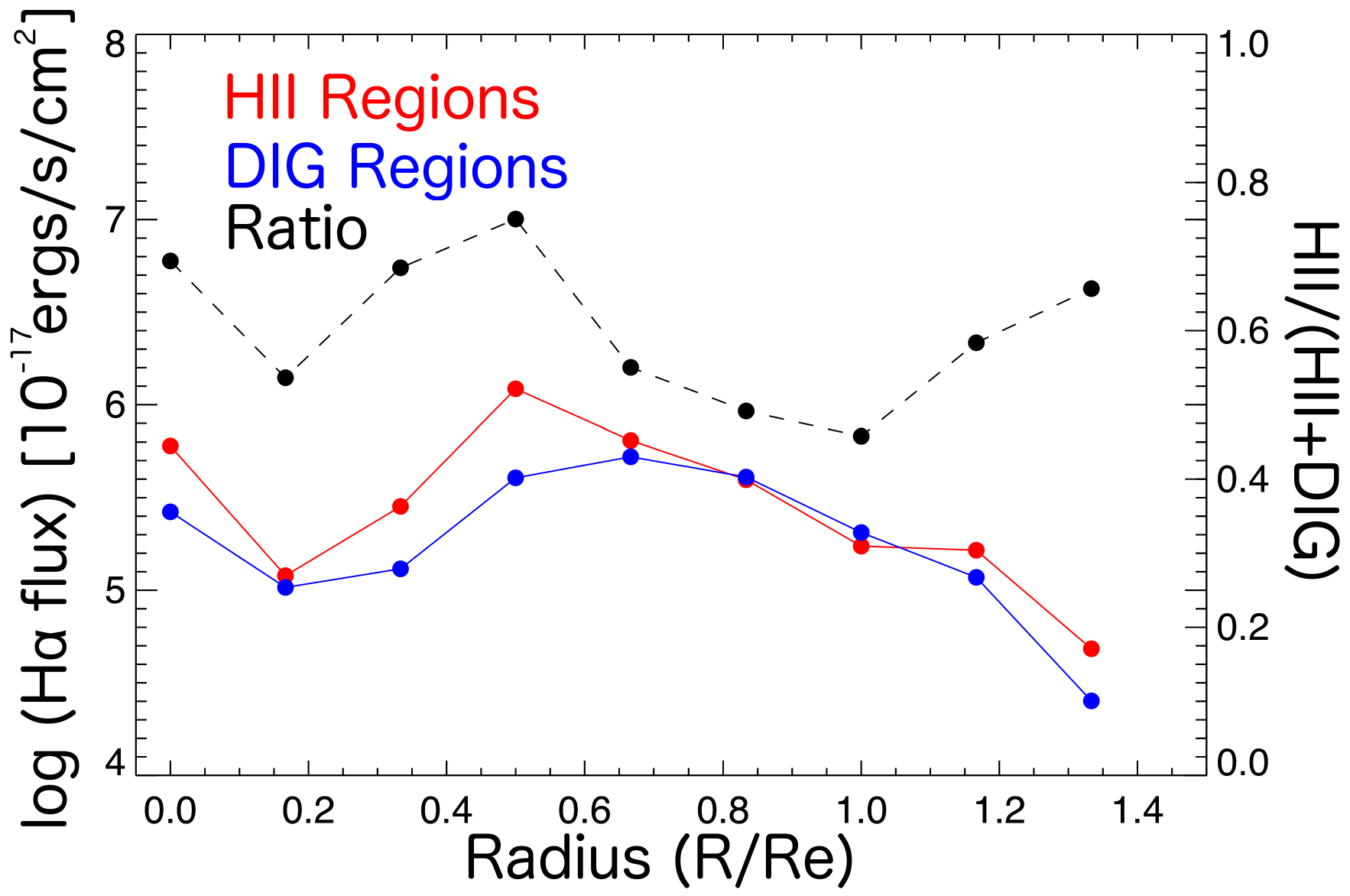}
	\caption{The H$\alpha$ luminosity radial profile of M83. The luminosity profiles of \htwo and DIG regions are almost identical meaning that the total H$\alpha$ luminosity is almost evenly contributed by \htwo regions and DIG regions. Even though \htwo regions are significantly brighter (Figure \ref{HASBProfile}), there are far more DIG regions (Figure \ref{number}) to contribute to the luminosity.}
	\label{luminosity}
\end{figure}
Using the [SII]/H$\alpha$ \htwo classification scheme, we analyze the basic properties of the DIG regions. Figure \ref{HASBProfile} shows the H$\alpha$ surface brightness radial profiles of \htwo regions and DIG regions for spaxels where S/N(H$\alpha$)$> 3$ and S/N(H$\beta$)$> 3$ at the native resolution. We note that the surface brightness profile shapes of the two regions are extremely similar, with the \htwo regions being consistently brighter than the DIG regions by about 0.6 dex. As expected, the peak of the H$\alpha$ surface brightness occurs at the centre. However, an increase in surface brightness of about 0.2 dex from the previous radial bin happens at R/R$_{e} = 0.6$ corresponding to the location of the spiral arms.
\par
Figure \ref{number} shows that there are more DIG spaxels than \htwo region spaxels at each radial bin, with the \htwo regions peaking at 45\% of the total spaxels at the location of the spiral arms. For the [SII]/H$\alpha$ classification scheme, just 12\% of spaxels are classified as \htwo regions (Table \ref{classification}). The number of spaxels for both \htwo and DIG regions steadily increase until we reach the spiral arms (R/R$_{e} = 0.6$). Beyond this radii, the number of detected spaxels begins to drop due to surface brightness limits.
\par
The combination of Figure \ref{HASBProfile} and Figure \ref{number} gives the H$\alpha$ luminosity profile. Figure \ref{luminosity} presents the H$\alpha$ luminosity profile for \htwo and DIG regions. The shape of these two profiles are very similar with only slight variations. Up to the spiral arms (R/R$_{e} = 0.6$), the total H$\alpha$ luminosity is slightly dominated by the \htwo regions. Beyond the spiral arms, the DIG regions contribute approximately half of the overall H$\alpha$ luminosity in radial annuli. 
\par
For each resolution scale, we create spatially resolved metallicity maps and their associated error maps using the metallicity diagnostics described in Section \ref{method}. Some of these metallicity diagnostics rely on the detection of emission lines that are relatively weaker than others. For example, \rtt and N2O2 rely on the accurate detection of the [OII] emission doublet which drastically reduces the amount of usable spaxels in their metallicity maps. Only 2674 (3\%) spaxels out of the total 72111 spaxels within 1.5 R$_{e}$ have [OII] emission with a S/N > 3. Metallicity diagnostics which use stronger emissions lines and do not require extinction correction such as N2S2 and N2HA have metallicity maps which cover a larger fraction of the galaxy, even in the areas dominated by DIG where the [OII] emission flux typically has S/N < 3. The [NII], [SII] and H$\alpha$ emission lines all have over 57000 (79\%) spaxels within 1.5R$_{e}$ with a S/N > 3.
\par
When determining the radial metallicity gradients at each resolution, we first correct for the observed ellipticity of the galaxy using the ellipticity and PA listed in Table \ref{galinfo}. We then use the robust line fitting program {\sc LTS$\_$LINEFIT} \citep{cappellari13} to fit a linear trend to the metallicity gradient at each resolution, propagating the uncertainty in metallicity through. The gradients are then normalized by the effective radius (R$_{e}$) of the galaxy to remove the size dependence of metallicity gradients \citep{sanchez14,ho15,sanchez-menguiano16}. \citet{sanchez-menguiano18} found wide-spread deviations from single linear metallicity gradients and instead uses multiple linear gradients to more accurately fit the radial metallicity distribution. For this study we adopt a single linear fit to our metallicity gradients.
\par
\begin{landscape}
	\begin{figure}
		\centering
		\includegraphics[width=1.3\textheight]{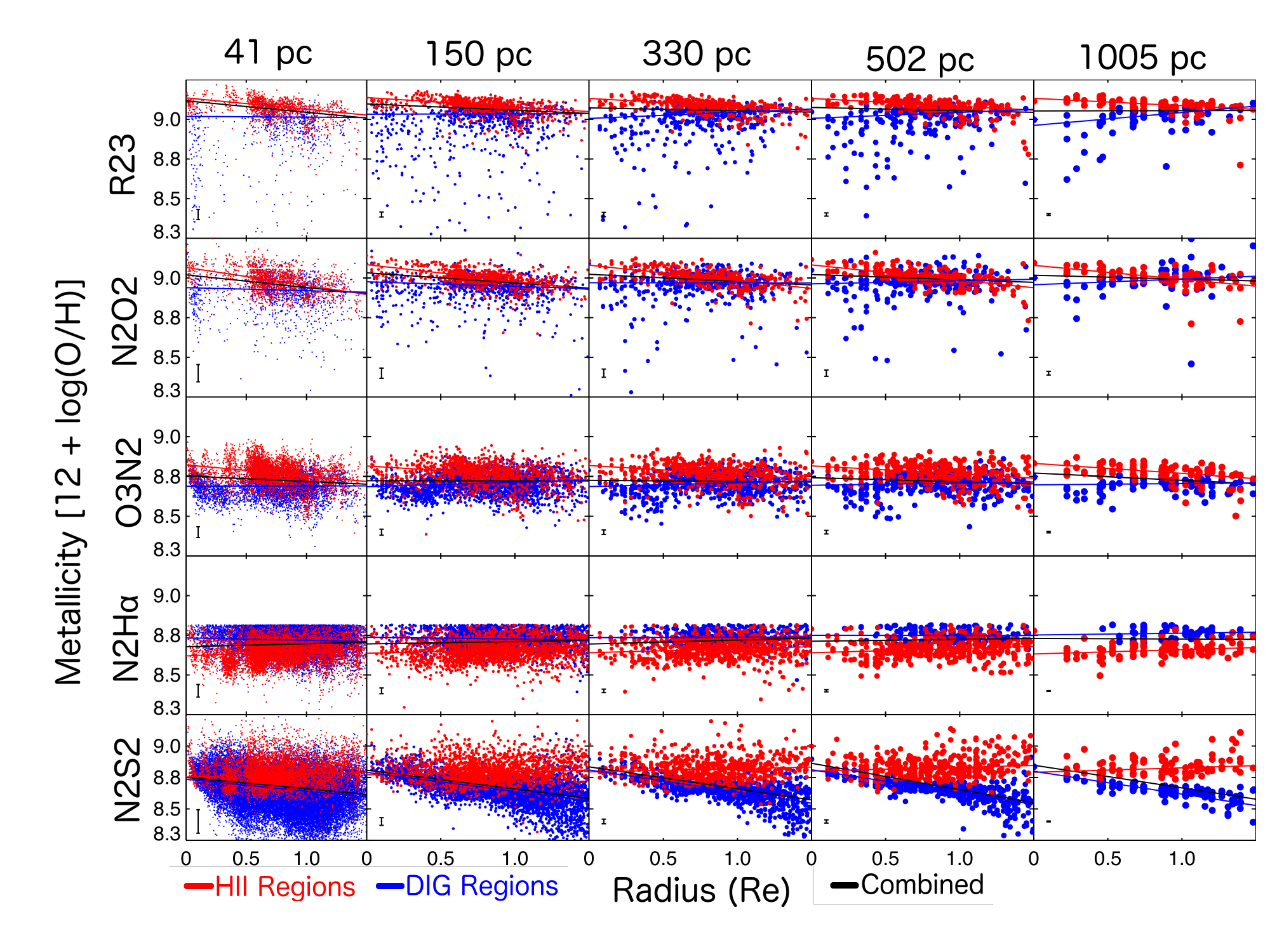}\label{fig:a}
		\caption{
Metallicity gradient linear fits for each metallicity diagnostic at each resolution scale. Red points represent \htwo regions and blue points represent the DIG regions. The black line shows the combined, non-separated metallicity gradients. We provide the median error of each panel in the bottom left corner. \htwo and DIG regions are separated using the \stha classification scheme.
		}
		\label{metgradgrid_siicut}
	\end{figure}
\end{landscape}

Figure \ref{metgradgrid_siicut} shows the measured metallicity gradients of M83 for the five chosen metallicity diagnostics at the five different spatial resolutions using the [SII]/H$\alpha$ DIG classification scheme. We show the same plots for the {\sc HIIphot} and H$\alpha$ surface brightness (SB) \htwo classification schemes in the supplementary material. In each panel we show the metallicity as a function of radius and fit the metallicity gradient using only \htwo regions (red), DIG regions (blue) and the full set (black) of spaxels. The median error of each panel is shown as black bars in the bottom left corner. Throughout this work we consider the radial metallicity gradient of the \htwo regions at the 41 pc to be the true metallicity gradient of M83. It is also important to note that the measured metallicity gradients of the DIG regions is non-physical and does not represent the true metallicity of the DIG. To measure the metallicity of the DIG, you need to apply a metallicity diagnostic which was specifically calibrated for DIG ionisation mechanisms \citep[e.g.][]{kumari19}. We simply apply the \htwo region calibrated diagnostics to the DIG regions to show how the emission line ratios of the DIG are handled by current emission line diagnostics. 
\par
 We summarise Figure \ref{metgradgrid_siicut} in Figure \ref{resplot} by plotting the metallicity gradient as a function of spatial resolution. The dashed red line represents the observed metallicity gradient when applying DIG corrections at the given resolution after contamination has already occurred. For example, the dashed red line in the \stha classification column at 502 pc represents the observed metallicity gradient when using combined emission spaxels at 502 pc and applying \htwo and DIG separation methods outlined in Section 3.1.1 and Section 3.1.2. At the native resolution of 41pc, the dashed red line is exactly the same as the red solid line. The original method outlined by \citet{kaplan16} requires a minimum of 200 spaxels in order to determine the characteristic \htwo and DIG [SII]/H$\alpha$ emission line ratio. When applying this method to the lower resolution scales with an insufficient spaxel count, we instead use brightest and dimmest 5\% of spaxels. We do not attempt to use {\sc HIIphot} at coarse resolution scales as it is optimised for high spatial resolution data and will likely fail at kiloparsec resolutions.
\par
The red line representing the change in \htwo region metallicity gradient with resolution demonstrates the effects of pure spatial smoothing. The black line demonstrates the effects of both spatial smoothing and DIG contamination. Figure \ref{resplot} shows that for the \rtt, N2O2 and O3N2 metallicity diagnostics, M83 possesses a negative metallicity gradient. However, for the N2H$\alpha$ and N2S2 metallicity diagnostics, M83 has a positive metallicity gradient. This highlights the significant differences in metallicity gradient calculations between different metallicity diagnostics.
\par
In general we see very similar trends from all the classification schemes across all metallicity diagnostics. The only outstanding difference occurs when applying a H$\alpha$ surface brightness cut to the N2H$\alpha$ metallicity diagnostic at 1005 pc. 
\subsection{The Impact of Different Metallicity Diagnostics}\label{metdiageffect}
\begin{landscape}
	\begin{figure}
		\centering
		\includegraphics[width=1.3\textheight]{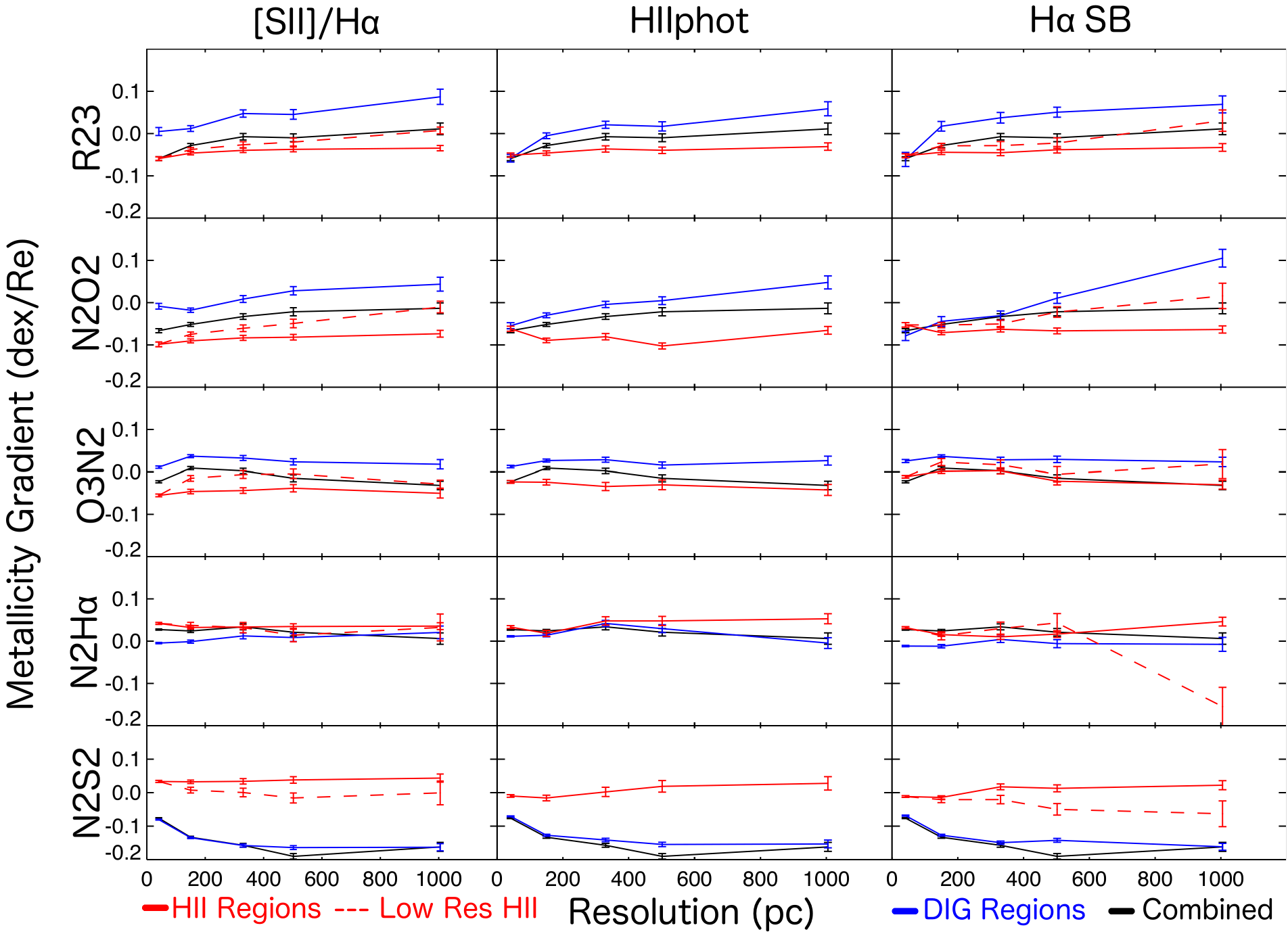}
		\caption{
We summarise the results of Figure \ref{metgradgrid_siicut} by plotting the metallicity gradients of each DIG classification scheme as a function of resolution scale. The red dashed line represents the metallicity gradient when the DIG classification scheme is applied after DIG contamination has occurred. Further detail is given in Section \ref{results}. We provide all values for metallicity gradient fits in Appendix A.
		}
		\label{resplot}
	\end{figure}
\end{landscape}

Figure \ref{resplot} show 3 different types of patterns. The metallicity gradient from both the DIG regions and full set of spaxels of the \rtt and N2O2 metallicity diagnostics tend to diverge away from the metallicity gradient of the \htwo regions, becoming shallower with coarser resolution. We expect spatial smoothing to flatten steeper gradients at a quicker rate. However, since the metallicity gradients of the combined spaxels and DIG regions appear to be flattening quicker than the steeper \htwo region metallicity gradients, another effect aside from DIG contamination must be present. 
\par
For the O3N2 metallicity diagnostic, the metallicity gradient of the full set of spaxels converge to the metallicity gradient of the \htwo regions as we move to coarser resolution scales, with agreement between the two at the 1005 pc scale. The metallicity gradient of the DIG regions tends to remain a relatively fixed offset ($\sim$0.075 dex/R$_{e}$) from that of the \htwo regions.
\par
The N2H$\alpha$ and N2S2 metallicity diagnostics appear to be the most impacted by DIG contamination. Except for the N2H$\alpha$ metallicity diagnostic with the H$\alpha$ DIG classification scheme, the metallicity gradient of the full sample appears to be tied to the DIG metallicity gradient. The N2S2 metallicity gradients produce significantly steeper metallicity gradients in the DIG regions compared to \htwo regions as opposed to shallower metallicity gradients in the other 3 metallicity diagnostics. The N2H$\alpha$ metallicity DIG gradients appear to be flat at all resolution scales.
\par
\begin{figure}
	\includegraphics[width=0.47\textwidth]{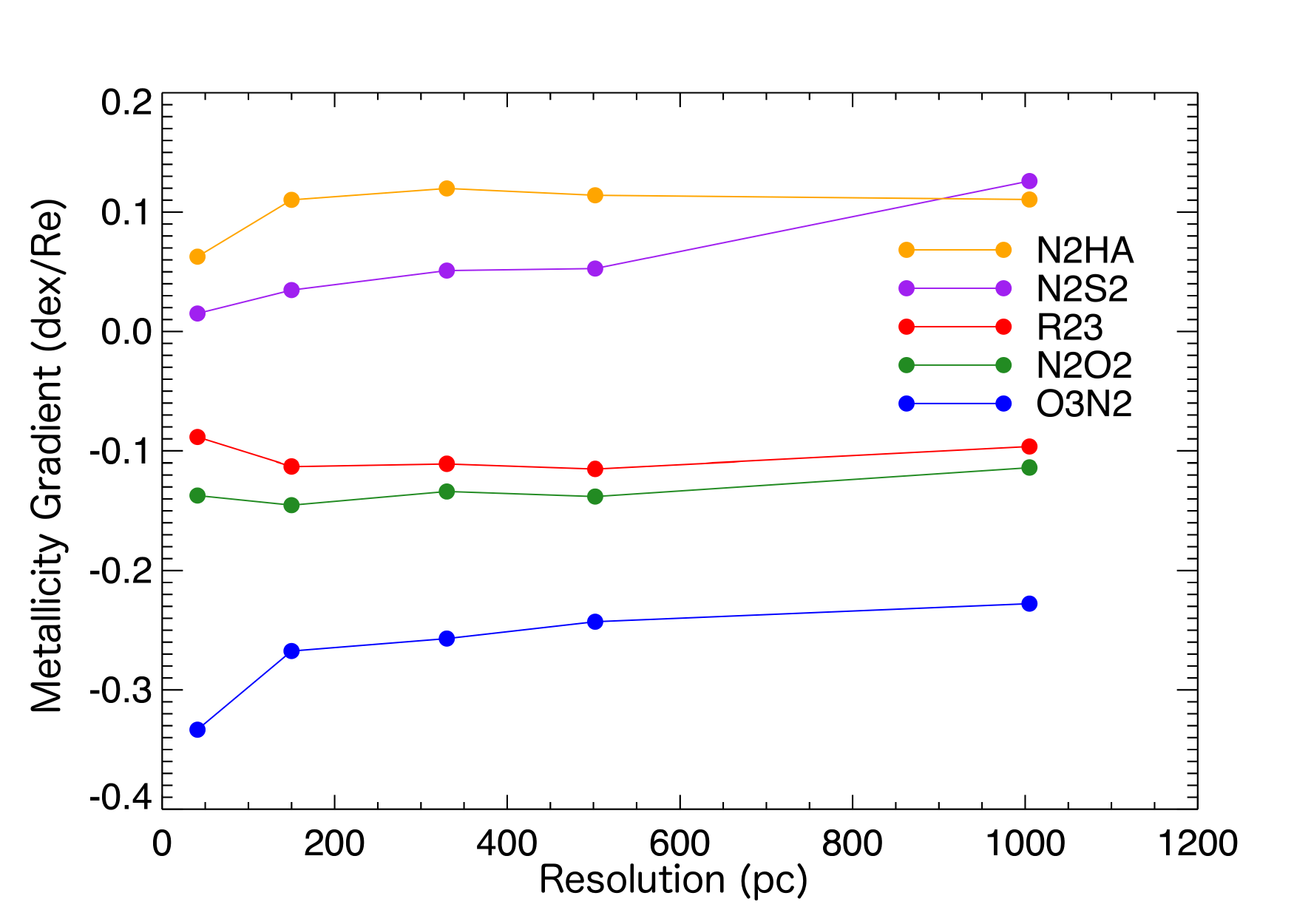}
	\caption{Metallicity gradients of \htwo regions where each diagnostic is restricted to using the same set of spaxels at each resolution scale. This highlights the significant systematic differences between the different metallicity diagnostics even when the same \htwo region spaxels are used.}
	\label{samespaxels}
\end{figure}

Although the systematic differences between the various metallicity diagnostics will not be discussed in detail in this paper, it is important to note that between different metallicity diagnostics, the metallicity gradient of \htwo regions do not agree with each other. Part of the differences seen between the \htwo metallicity gradients seen in Figure \ref{resplot} is caused by the different set of spaxels available to each diagnostic. Figure \ref{samespaxels} shows the metallicity gradients of \htwo regions where the same set of spaxels are used for all metallicity diagnostics. Figure \ref{samespaxels} clearly demonstrates the lingering issues that exist between the different metallicity diagnostics. Using the Sloan Digital Sky Survey (SDSS), \citet{kewley08} provides empirical conversions for a suite of metallicity diagnostics and calibrations for aperture metallicities. Until a similar study has been done for metallicity gradients, metallicity gradients determined from different metallicity diagnostics should not be compared to one another.


\section{Discussion}
\subsection{Low Surface Brightness Regions}\label{LSBdiscussion}
Spatial smoothing and diffuse ionized gas (DIG) contamination are not the only factors which affect the metallicity gradient when using coarser resolutions. Metallicity diagnostics that use weak or low S/N emission lines have fewer spaxels with which to determine the metallicity gradients. As the data is smoothed, the S/N increases, meaning spaxels containing lower surface brightness emission are now included in the gradient determination. 
\par
In our data, metallicity diagnostics that use [OII] such as \rtt and N2O2 are more impacted by low surface brightness (LSB) regions than other diagnostics. Furthermore, as each diagnostic uses different sets of emission lines, they will also have different sets of LSB spaxels and will be affected differently by their inclusion. Different \htwo region classification schemes also contribute to differences in the set of spaxels used. Figure \ref{resplot} demonstrates that the general trends are preserved regardless of which classification is used, so we therefore select the [SII]/H$\alpha$ classification scheme as our fiducial model for the following discussion.
\par
In this paper, we do not separate the effects of spatial smoothing and the inclusion of LSB \htwo region spaxels. We briefly discuss the effects of LSB regions in Section \ref{spatialsmoothing}.
\begin{table}
	\centering
	\begin{tabular}{c|ccc}
	\multicolumn{4}{c}{\htwo spaxels} \\
	Diagnostic & [SII]/H$\alpha$ & HIIPhot & H$\alpha$ SB \\
	\hline
\rtt &     0.14&     0.25&     0.15 \\
N2O2 &     0.17&     0.28&     0.18 \\
O3N2 &     0.48&     0.74&     0.56 \\
N2H$\alpha$ &     0.86&     0.93&     0.90 \\
N2S2 &     0.86&    1.00&      1.00 \\
\end{tabular}
\caption{
Fraction of \htwo region spaxels with metallicity estimates at 41 pc. As the metallicity diagnostics require less emission lines or use stronger detected emission lines, the fraction increases. These fractions are relative to the numbers presented in Table \ref{htwodignumbers}.
}
\label{ratioHSB}
\end{table}
\subsection{Spatial Smoothing}\label{spatialsmoothing}
Several factors are believed to be contributing to metallicity gradient flattening, including the effect of spatial smoothing. Spatial smoothing is the effect of averaging regions of high and low metallicity together when using larger spaxel sizes. The effect of spatial smoothing is most prominent with steep gradients and obviously does not affect galaxies that possess flat gradients \citep[e.g.~NGC\,5474][]{mast14}. 
\par
The change in metallicity gradient with resolution in the pure \htwo region data demonstrates this effect of spatial smoothing. Overall in Figure \ref{resplot} we can see that \htwo region metallicity gradients generally remain constant, with all the \htwo metallicity gradients remaining within $\pm0.01$ dex for the [SII]/H$\alpha$ classification scheme and within $\pm0.02$ dex for {\sc HIIphot} and the H$\alpha$ classification scheme.
\par
As mentioned in Section \ref{LSBdiscussion}, LSB \htwo regions (where e.g. [OII] is not of sufficient S/N), are also impacting the change in the \htwo region metallicity gradient with resolution. Table \ref{ratioHSB} and \ref{ratioHSB_dig} reveal that the \rtt and N2O2 metallicity diagnostics are only possible in a very small fraction of spaxels. This is mostly driven by the lower S/N of [OII], which is due to the higher noise in the blue part of the TYPHOON spectra. We therefore expect \rtt and N2O2 to be the most heavily affected by the effects of the inclusion of LSB regions as we decrease the resolution. The \rtt and N2O2 \htwo region metallicity gradients are flattened relatively more compared to those calculated using the N2H$\alpha$ and N2S2 metallicity diagnostics. However, this is also consistent with the fact that \rtt and N2O2 have steeper gradients to begin with and are therefore naturally more affected by pure spatial smoothing.
\par
For the following sections, we emphasise once again that these metallicity diagnostics can not be applied to emission originating from the DIG. However, we apply them to the DIG regions in order to assist the reader in understanding the effects of DIG contamination and refer to them as "DIG metallicity gradients" for simplicity. The DIG metallicity gradients are clearly affected by more than just spatial smoothing. The DIG metallicity gradients from the N2S2 metallicity diagnostic become steeper (-0.095 dex/R$_{e}$ at 41 pc to -0.181 dex/R$_{e}$ at 1005 pc with the [SII]/H$\alpha$ classification scheme) as we decrease the resolution. Whereas the DIG metallicity gradients from the \rtt and N2O2 metallicity diagnostics start off with a negative gradient (-0.005 and -0.019 dex/R$_{e}$ respectively with the [SII]/H$\alpha$ classification scheme) and are positive at the 1005 pc resolution (0.081 and 0.036 dex/R$_{e}$ respectively with the [SII]/H$\alpha$ classification scheme). The O3N2 and N2H$\alpha$ DIG metallicity gradients appear to remain fairly constant and relatively unaffected (within $\pm0.02$ dex/R$_{e}$) by the decreasing resolution. The steepening of the negative DIG metallicity gradient of N2S2 and transition from negative to positive metallicity gradient of the R23 and N2O2 diagnostics can not be explained by spatial smoothing alone.
\par
Table \ref{ratioHSB_dig} shows the ratio of DIG spaxels with metallicity measurements. The \rtt and N2O2 metallicity diagnostics at a resolution scale of 41 pc are measuring the metallicity gradient of the DIG using only 1\% and 2\% of the total amount of DIG spaxels respectively. As we increase the spaxel size, we gradually include more LSB DIG regions which causes the DIG metallicity gradient to steepen towards the positive direction. As we are only starting with 1\% and 2\% of DIG spaxels, the DIG metallicity gradient continues to steepen all the way to the coarsest resolution.
\par
For the N2S2 metallicity diagnostic, we are measuring the DIG metallicity gradient with 68\% of DIG spaxels at a resolution scale of 41 pc. Since most of the DIG spaxels are already included at the finest resolution, less binning is needed to include all the emission from the LSB DIG regions. At a resolution of about 330 pc, the N2S2 DIG metallicity gradients stop steepening and remain constant within the uncertainties for coarser resolution scales. This is likely due to all the LSB DIG spaxels being included at the 330 pc scale and pure spatial smoothing happening as we continue to degrade the resolution. 
\begin{table}
	\centering
	\begin{tabular}{c|ccc}
		\multicolumn{4}{c}{DIG spaxels} \\
		Diagnostic & [SII]/H$\alpha$ & HIIPhot & H$\alpha$ SB \\
		\hline
\rtt &    0.01&    0.01&    0.01 \\
N2O2 &    0.02&    0.02&    0.01 \\
O3N2 &    0.07&    0.09&    0.04 \\
N2H$\alpha$ &     0.29&     0.32&     0.26 \\
N2S2 &     0.68&     0.68&     0.65 \\
	\end{tabular}
	\caption{
Same as Table \ref{ratioHSB} for DIG spaxels.
	}
	\label{ratioHSB_dig}
\end{table}
Although the O3N2 metallicity diagnostic only uses 7\% of DIG spaxels at a resolution scale of 41 pc, the DIG metallicity gradients remain within $\pm0.02$ dex/R$_{e}$ at all resolution scales. This suggests that the ([OIII]/H$\beta$)/([NII]/H$\alpha$) emission line ratio is relatively independent of H$\alpha$ surface brightness, and hence the inclusion of low surface brightness DIG spaxels has no effect on the emission line ratio radial profile.
\par
It is difficult to assess how the N2H$\alpha$ DIG metallicity diagnostic evolves with varying spatial resolution. We expect the [NII]/H$\alpha$ ratio to increase in DIG regions, and hence the N2H$\alpha$ metallicity. However, the PP04 N2H$\alpha$ metallicity diagnostic imposes an upper and lower limit on the \ntha line ratio. Most of the DIG spaxels lie above this upper limit, meaning that most of the DIG is removed from the metallictity gradient measurements. This upper limit can be seen clearly in Figure \ref{metgradgrid_siicut}. Table \ref{ratioHSB_dig} shows that only $\sim$29\% of DIG spaxels lie within the limits of this diagnostic.
\subsection{DIG Contamination of \htwo Regions}
At increasingly coarse resolution scales, the boundary between \htwo regions and the DIG becomes unclear, leading to an increasing DIG contamination in the \htwo region emission line spectrum. DIG dominated regions have different ionization mechanisms and tend to have increased line ratios ([SII]/H$\alpha$, [NII]/H$\alpha$) compared to \htwo regions \citep{blanc09,zhang17}. The inclusion of DIG therefore systematically alters the line ratios of affected spaxels, leading to changes in their metallicity measurements. 
\par
As discussed in Section \ref{metdiageffect}, Figure \ref{resplot} shows that the \rtt and N2O2 metallicity diagnostic gradients of the full set of spaxels diverges away from those determined using only pure \htwo region spaxels. If metallicity gradient flattening for the full set of spaxels was solely due to spatial smoothing, we would be expecting these two quantities to converge since spatial smoothing has a bigger effect on steeper gradients.
\par
\begin{figure}
\centering
\includegraphics[width=0.5\textwidth]{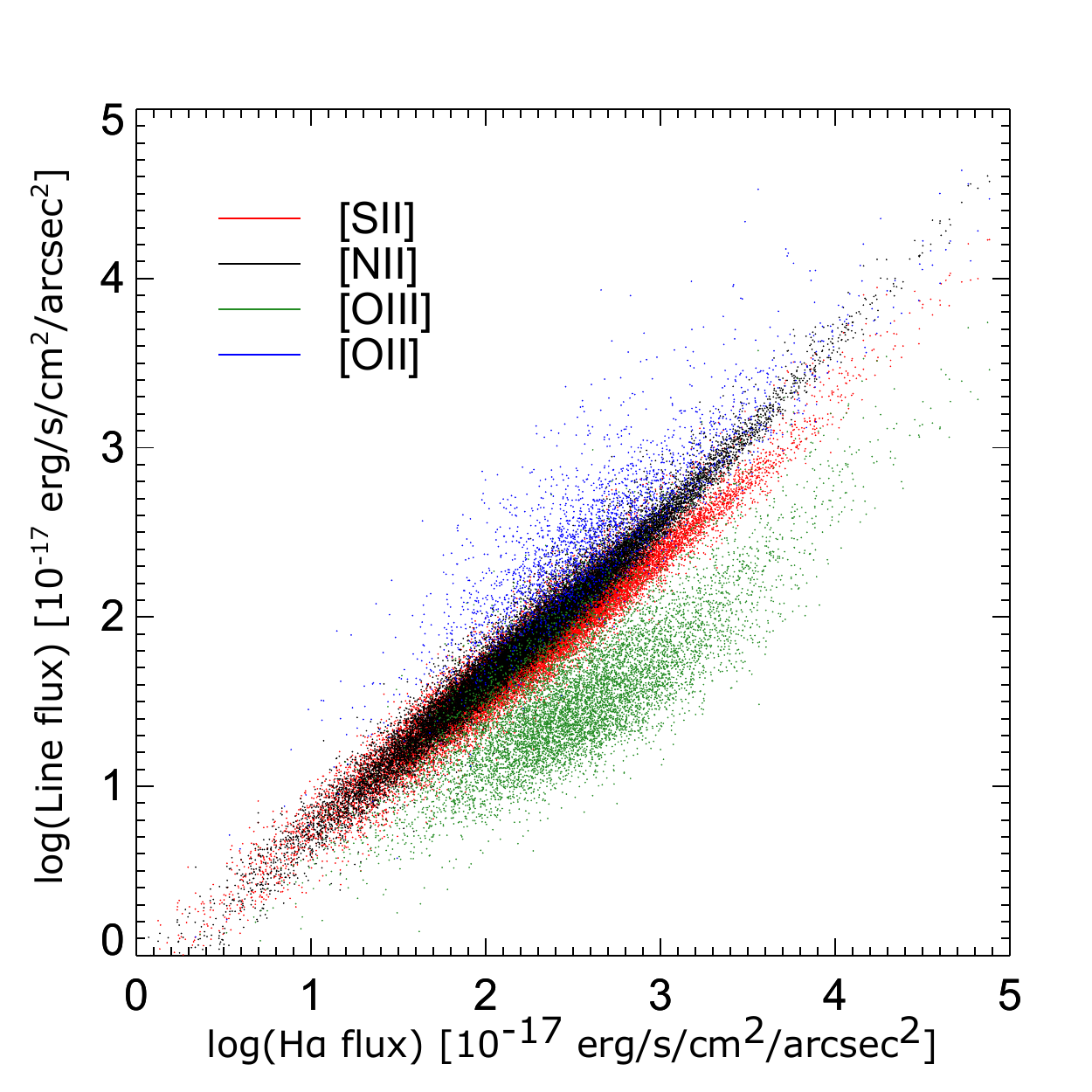}
\caption{
An almost linear relationship is seen between the H$\alpha$ flux and all other strong emission line fluxes. This means that H$\alpha$ can be used as a proxy for the strength of all other strong emission lines. [OII] and [OIII] have the greatest scatter from a linear relationship but can still be sufficiently described by the H$\alpha$ flux.
}
\label{havsline}
\end{figure}

Figure \ref{havsline} shows the relationship between the flux of various strong emission lines and H$\alpha$. An almost linear relationship is seen amongst all the strong emission lines. This means that when spaxels are merged together at coarser resolution scales, the resulting emission line ratio is a luminosity weighting of the line ratios of the individual spaxels. We can therefore determine how much of an impact DIG contamination has on emission line ratios by looking at the amount of H$\alpha$ flux contributed by the DIG. We determine the \htwo dominance in each coarse-resolution spaxel by taking the ratio of the H$\alpha$ surface brightness of the \htwo region datacube relative to that in the combined emission datacube. We then take the median of each radial bin to create a radial profile of the \htwo dominance. As each metallicity gradient is measured using a different set of spaxels, we create a \htwo region dominance profile for each metallicity diagnostic at every resolution scale above native (41 pc). This shows how each metallicity diagnostic is affected differently by DIG contamination. 
\par

\begin{landscape}
	\begin{figure}
		\centering
		\includegraphics[width=1.3\textheight]{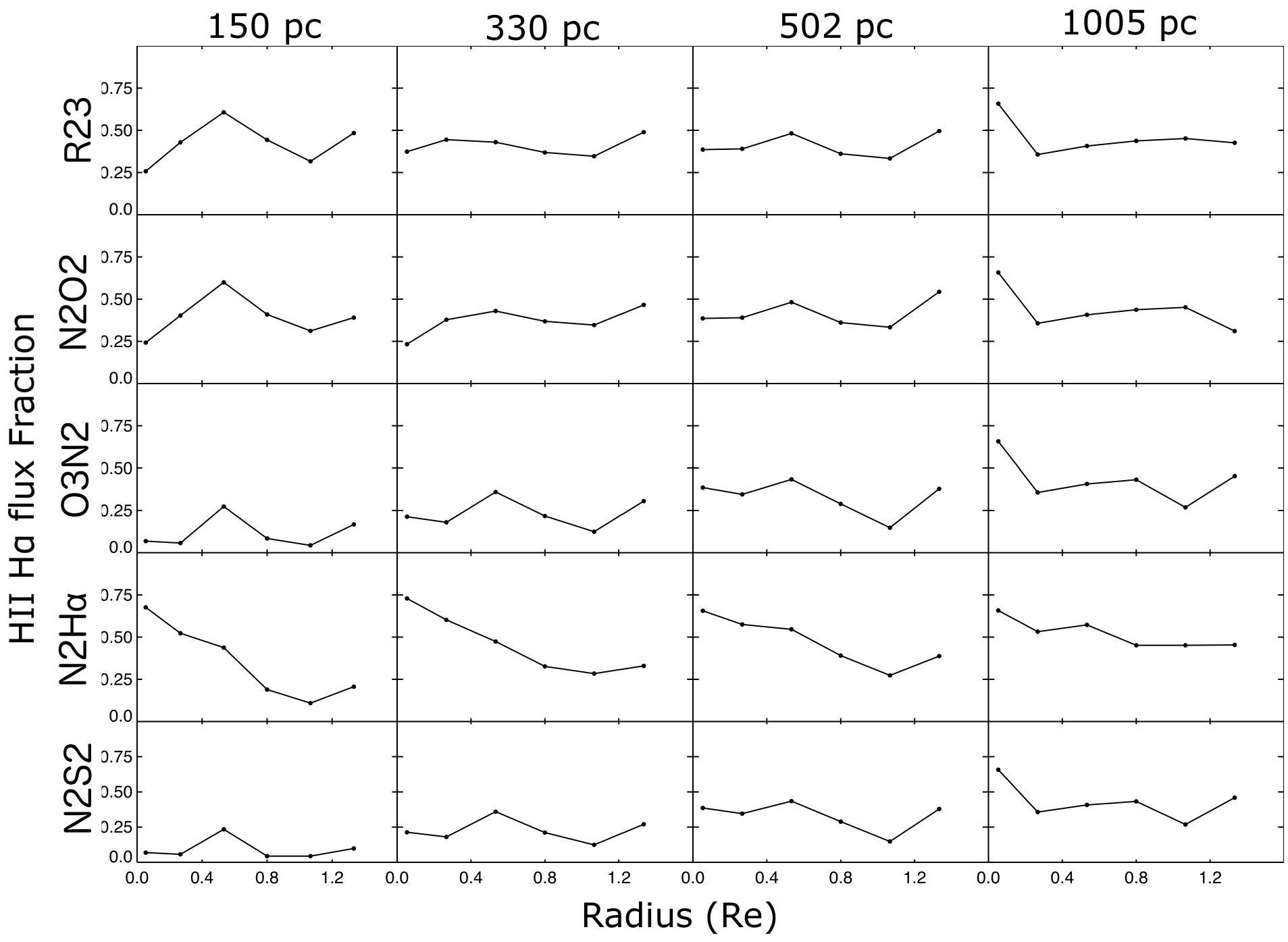}\label{fig:a}
		\caption{
 Here we show the median fraction of flux originating from \htwo regions in radial bins of 0.25R$_{e}$ for each metallicity diagnostic at each binned resolution based on the spaxels used in each metallicity gradient fit. Since each metallicity gradient fit uses a different set of spaxels, the \htwo dominance profiles are different for each metallicity diagnostic. The difference in \htwo dominance profiles is dictated by which emission lines are used in the metallicity diagnostics. \rtt and N2O2 have essentially identical profiles due to the limitations imposed by the [OII] emission line. At a resolution of 1005 pc, all metallicity diagnostics are basically using the same spaxels, hence the \htwo dominance profiles are very similar. 
		}
		\label{hiidominance}
	\end{figure}
\end{landscape}

From Figure \ref{hiidominance}, we can see that DIG contamination affects each diagnostic differently and the \htwo dominance profile is constantly changing at each resolution. As the resolution is decreased, the differences between the \htwo dominance profiles begin to disappear since at the coarsest resolution, the set of spaxels used for each metallicity diagnostic are essentially the same.
\par
The \htwo dominance profiles of \rtt and N2O2 are the same to within 10\% of each other. This is expected as both diagnostics are both limited by the [OII] emission line. At a resolution scale of 150 pc, the \htwo dominance profile has a spike at approximately R/R$_{e} = 0.6$ corresponding with the location of the spiral arms. At the resolution scales of 330 and 502 pc, the \htwo dominance profile consistently stays within 30\% - 50\%. 
\par
The inner-most radial bin ($0 <$ R/R$_{e} < 0.25$) for \rtt and N2O2 demonstrates the transition from being dominated by sheer number of DIG spaxels to luminosity weighting at 1005 pc. When fitting the metallicity gradients, \htwo region and DIG region spaxels are numerically weighted rather than luminosity weighted, meaning that a large number of DIG spaxels is able to overcome a small number of higher luminosity \htwo region spaxels. However, the merging of spaxels is a luminosity weighted action. As there are a relatively small number of DIG spaxels compared to their surface brightness in the inner radii, the higher surface brightness \htwo regions dominate in this process. However, in the outer regions, the \htwo dominance remains stable within 30\% - 50\% . This is because in the outer regions, there exists sufficient DIG spaxels to overcome the higher surface brightness \htwo regions.
\par
The O3N2 and N2S2 \htwo dominance profiles demonstrates the transition from numeric weighting to luminosity weighting at all radii rather than just the inner radii as seen in \rtt and N2O2. Since the O3N2 and N2S2 diagnostics use very strong emitting lines (no [OII] dependence), at fine resolution scales, the large number of DIG spaxels numerically dominate metallicity gradient determinations over the fewer more luminous \htwo regions. As we move to coarser resolution scales, luminous \htwo regions dominate the emission in increasingly more spaxels until the final \htwo dominance profile at 1005 pc resembles the other diagnostics.
\par
While the DIG remains within the calibration limits of the \rtt, N2O2, O3N2 and N2S2 diagnostics, the same is not true for the N2H$\alpha$ diagnostic. Since the DIG has a significantly enhanced \ntha ratio, the upper limit removes most of these spaxels from the metallicity gradient calculation. Due to the inherent nature of the \ntha calibration, the weighting effect caused by the sheer number of DIG spaxels is severely reduced. A negative gradient \htwo dominance profile is persistent throughout every resolution scale. 
\par
Figure \ref{lineprofile} shows the changes in the emission line ratio profiles of M83 at the coarse resolution scales. As expected, the emission line ratio profiles of \htwo regions are relatively unchanged as we decrease the resolution across all diagnostics. The emission line profiles of the DIG regions change significantly with decreasing resolution in the ([OII]+[OIII])/H$\beta$ and [NII]/[OII] emission line ratios. This is a result of the inclusion of emission from LSB regions at coarser resolution scales. However, from Table \ref{ratioHSB_dig} we see that the ([OIII]/H$\beta$)/([NII]/H$\alpha$) emission line ratio should also be significantly affected by the inclusion of LSB regions as very few DIG spaxels are present at the native resolution. The ([OIII]/H$\beta$)/([NII]/H$\alpha$) emission line ratio profile remains relatively unchanged across all resolution scales despite the inclusion of LSB regions. This suggests that LSB DIG regions have similar ([OIII]/H$\beta$)/([NII]/H$\alpha$) emission line ratios to the high surface brightness (HSB) DIG regions, while being significantly different for the ([OII]+[OIII])/H$\beta$ and [NII]/[OII] emission line ratios.
\par
For the ([OII]+[OIII])/H$\beta$ and [NII]/[OII] emission line ratio profile, we can see a clear difference between the profiles of the \htwo and DIG regions. For both of these emission line ratios, the \htwo regions and DIG regions exhibit opposite radial gradients at the coarsest resolution scale. The upper-branch (12+log(O/H)>8.5) of the \rtt metallicity diagnostic is approximately a one-to-one anti-correlated function of the ([OII]+[OIII])/H$\beta$ emission line ratio. From Figure \ref{lineprofile} we can easily see that the \htwo regions have a positive radial gradient in the ([OII]+[OIII])/H$\beta$ emission line ratio, leading to a negative metallicity gradient, while the DIG has a negative radial gradient in the ([OII]+[OIII])/H$\beta$ emission line ratio, causing the observed positive metallicity gradient in the DIG regions. Conversely, the N2O2 metallicity diagnostics is a one-to-one positively correlated function of the [NII]/[OII] emission line ratio. We see that for \htwo regions, the [NII]/[OII] emission line ratio has a negative radial gradient, which produces the negative metallicity gradient. The DIG regions have a positive radial gradient for the [NII]/[OII] emission line ratio, leading to the positive metallicity gradient in the DIG regions.
\par
By combining Figure \ref{hiidominance} and Figure \ref{lineprofile}, we can get an understanding of how the DIG can severely affect the measured metallicity gradient at low resolution scales where \htwo and DIG emission can not be separated. The dominance profiles for \rtt and N2O2 are approximately constant at around 0.4 and show a negative radial gradient at the coarsest resolution scale. If we combine the \htwo region and DIG region emission line ratio profiles together using Figure \ref{hiidominance} as weights, we can see that the metallicity gradient would be flattened as a result. With \htwo regions and DIG regions having opposite emission line ratio gradients and the constant \htwo dominance profile at 0.4, the \htwo region and DIG regions cancel each other out, leaving a flat emission line ratio profile and hence no metallicity gradients when combined.
\par
Applying this same process to the other metallicity diagnostics explains most of the behavior that we see in Figure \ref{resplot}. The O3N2 metallicity diagnostic is a negative linear fit to the ([OIII]/H$\beta$)/([NII]/H$\alpha$) emission line ratio. The O3N2 \htwo dominance profile shows us that the metallicity gradient fit is initially dominated by DIG, with the \htwo regions slowly taking over in the inner-sections of the galaxy as we decrease the resolution. Neither the \htwo region or DIG region emission line ratio profiles change significantly as a function of resolution. This means that as the inner-sections of the galaxy tend towards the \htwo regions while the outer regions remain the same, we end up with a steepening of the metallicity gradient as we decrease the resolution scale.
\par
The [NII]/H$\alpha$ emission line ratio profiles for \htwo regions and DIG regions undergo no changes as we decrease the resolution. This is due to the upper limit of the N2H$\alpha$ metallicity diagnostic which places boundaries on [NII]/H$\alpha$ for which the diagnostic can be used. The N2H$\alpha$ metallicity diagnostic is a one-to-one positively correlated function with the [NII]/H$\alpha$ emission line ratio. There is a very small positive gradient in the \htwo region [NII]/H$\alpha$ emission line ratio profile which causes the positive metallicity gradient. The [NII]/H$\alpha$ emission line profile for the DIG regions shows that its lowest emission line ratio occurs at [NII]/H$\alpha = -0.3$. This coincidentally corresponds to the upper limit of the N2H$\alpha$ metallicity diagnostic which can only be used for $-2.5 < $[NII]/H$\alpha < -0.3$. This explains why the DIG metallicity gradients are flat for the N2H$\alpha$ metallicity diagnostic at all resolution scales. As the resolution is decreased, the [NII]/H$\alpha$ ratio can only increase from its lowest point of [NII]/H$\alpha = -0.3$. Since anything above [NII]/H$\alpha > -0.3$ is simply cut by the boundaries of the calibration, nothing but a flat gradient can exist.
\par
Since the \htwo regions are also positioned quite close to the upper limit ([NII]/H$\alpha = -0.4$), as the resolution is decreased, the combined emission also exceeds this upper limit, causing the flat metallicity gradients. Based off of the \htwo region dominance profiles of the other 4 metallicity diagnostics and the emission line ratio radial profile of [NII]/H$\alpha$, we would expect to see a positive metallicity gradient at low resolution scales for lower metallicity galaxies where we do not exceed the upper limit. The high \htwo region dominance in the inner-region combined with the increased DIG dominance and increasing [NII]/$H\alpha$ ratio in the outer regions lead to an overall positive [NII]/H$\alpha$ emission line ratio radial profile.
\par
Much like the [NII]/H$\alpha$ emission line ratio, the $\log$([NII]/[SII]) + 0.264$\times\log$([NII]/H$\alpha$) emission line ratio also does not change significantly as we decrease the resolution scale. This is expected since the majority of spaxels already have a metallicity estimate and thus the impact of including emission from LSB regions is minimal. The N2S2 metallicity diagnostic is very closely approximated by a linear fit with a gradient of 1. The \htwo dominance profile of the N2S2 metallicity diagnostic makes it very clear why the combined metallicity gradient is almost perfectly tied with the DIG metallicity gradient. Similar to the O3N2 metallicity diagnostic, the \htwo dominance profile of N2S2 starts off with a very low \htwo dominance at all radii with the inner-section becoming more \htwo dominated relative to the outer radii as we decrease the resolution. We can clearly see that the \htwo regions have a higher $\log$([NII]/[SII]) + 0.264$\times\log$([NII]/H$\alpha$) emission line ratio at all radii compared to the DIG, meaning a higher absolute metallicity overall. The flat radial distribution of the \htwo regions means that it does not compensate for the DIG emission like the \rtt and N2O2 metallicity diagnostics, meaning the combined emission metallicity gradients are dictated by the DIG emission.
\subsection{Implications for Low Resolution Observations}
As we have thoroughly demonstrated, measurements of the radial metallicity gradient deviate significantly from the true metallicity gradient as we decrease the spatial resolution. We have shown that contamination by the DIG and LSB regions have a far greater effect on the metallicity gradient than spatial smoothing. This means that the removal of DIG is an extremely important step in determining accurate radial metallicity gradients.
\par
Ideally, the best method of obtaining the true metallicity gradient is to isolate the pure \htwo regions, however this is not always possible at higher redshift due to the limits of angular resolution. Therefore, we require either a metallicity diagnostic that is less sensitive to the DIG or some way to estimate and correct for the DIG contamination at the resolution of our observations. 
\par
The dashed red lines in Figure \ref{resplot} show how each \htwo classification scheme performs at coarser resolution scales. In most cases, attempting to correct for DIG after contamination has occurred, results in metallicity gradients closer to their true value, with benefits diminishing as the resolution is decreased. For the [SII]/H$\alpha$ \htwo classification scheme, applying the DIG correction either improves metallicity gradient measurements or has no effect for all metallicity diagnostics and resolution scales used. For the N2O2, N2HA and O3N2 metallicity diagnostic using the H$\alpha$ surface brightness \htwo classification scheme, applying a DIG correction at 1005 pc results in a worse measurement of the metallicity gradient.
\par
\begin{landscape}
	\begin{figure}
		\centering
		\includegraphics[width=1.3\textheight]{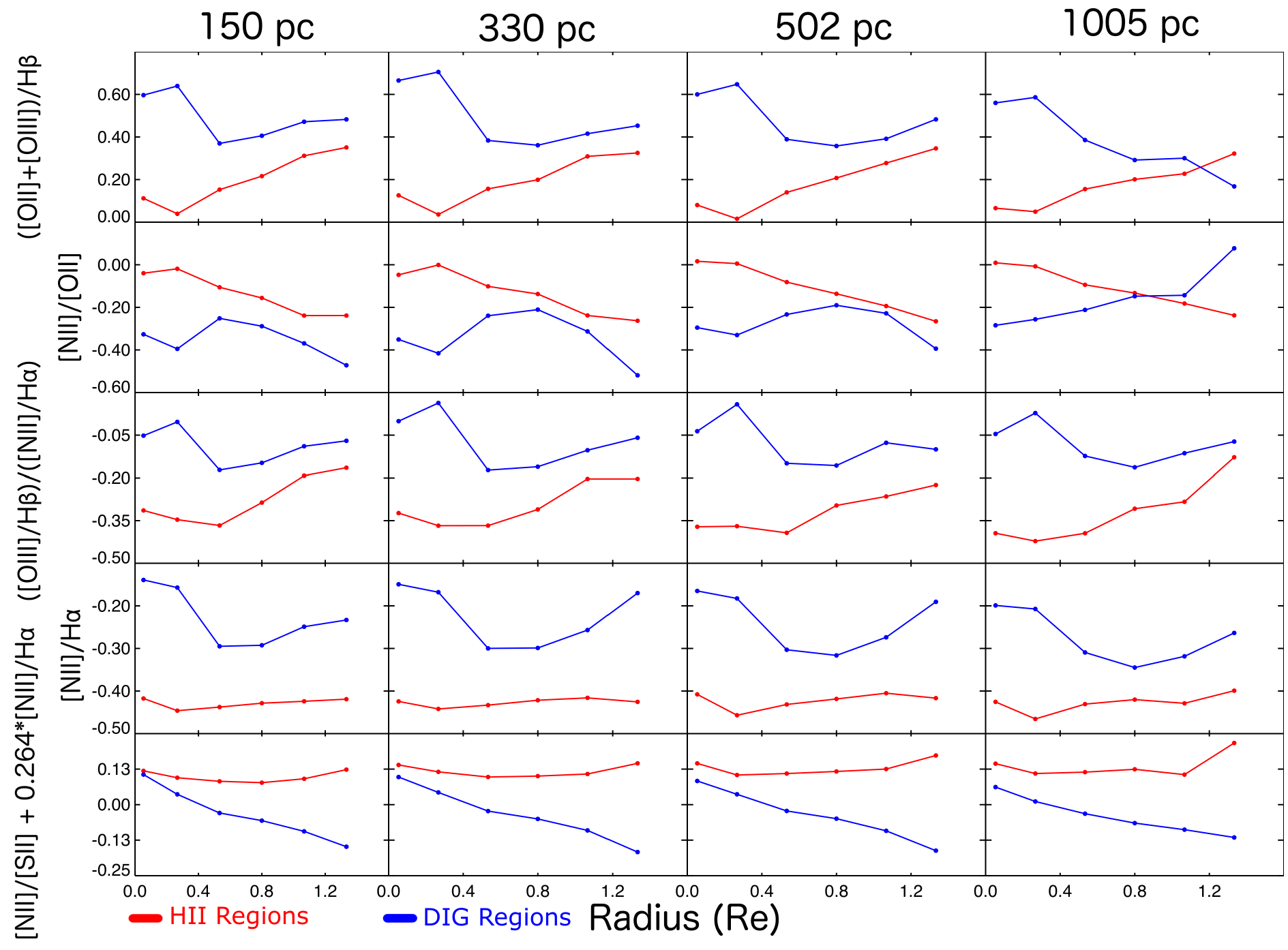}\label{fig:a}
		\caption{
The emission line ratio profiles of each metallicity diagnostic at each binned resolutions in radial bins of 0.25R$_{e}$. The \htwo region emission line ratios are largely unaffected by decreasing the resolution scales. However, the DIG emission line ratios are significantly different depending on how coarse the resolution scale is. The change in the DIG emission line profiles are caused by the inclusion of LSB regions as the data is binned.
		}
		\label{lineprofile}
	\end{figure}
\end{landscape}

Figure \ref{lineprofile} shows that the radial emission line ratio profiles of the DIG is often in the opposite direction to the \htwo region profiles. Any amount of DIG contamination would lead to shallower metallicity gradients. Applying the \htwo classification scheme to contaminated data removes all of the 'pure' DIG spaxels, but is unable to remove spaxels that contain a mixture of \htwo and DIG regions. By removing the 'pure' DIG spaxels, we are improving the measurement of the metallicity gradient but are unable to remove the effects of DIG completely. 

\section{Summary}
We have presented a systematic study of how the diffuse ionized gas (DIG) affects the metallicity gradient determined from a range of metallicity diagnostics using high spatial resolution (41 pc) IFS data on the face-on spiral M83. We separate the \htwo and DIG regions using three different \htwo region classification schemes: the [SII]/H$\alpha$ ratio, {\sc HIIphot} \citep{thilker00} and the H$\alpha$ surface brightness. Once the regions have been separated, we rebin the datacubes to 150, 330, 502 and 1005 pc to simulate low resolution observations.
\par
We find that the average H$\alpha$ surface brightness of the \htwo regions is about 0.6 dex greater than the DIG regions at a given radii for R/R$_{e} < 1.5$. However, depending on which \htwo classification scheme is used, DIG and low surface brightness spaxels comprise between 85\% - 94\% of spaxels within R/R$_{e} < 1.5$. Although \htwo regions are significantly brighter than the DIG regions, DIG region spaxels are vastly more numerous. These two effects mean that the total H$\alpha$ luminosity is approximately contributed by 70\% \htwo region emission and 30\% DIG emission for R/R$_{e} < 1.5$ in M83. The spiral arms of M83 extend to about R/R$_{e} = 0.6$, where the H$\alpha$ luminosity is 75\% contributed from \htwo regions. Beyond this ($0.6 <$ R/R$_{e} < 1.5$), \htwo regions contribute only 60\% of the total H$\alpha$ luminosity.
\par
Using five different metallicity diagnostics we calculate the radial metallicity gradients as a function of resolution and \htwo region classification scheme. We find that the inclusion of DIG significantly affects the measured metallicity gradient by more than we would expect from pure spatial smoothing. Using a [SII]/H$\alpha$ DIG classification scheme, the \rtt \citep{kobulnicky04} and N2O2 \citep{kewley02} metallicity gradients are significantly flattened by the DIG by up to 0.048 dex/Re and 0.063 dex/Re respectively. The O3N2 \citep{pettini04} metallicity gradients are initially flattened by the DIG by 0.057 dex/Re at 150 pc, with the metallicity gradients steepening and converging to within 0.02 dex/Re of the true metallicity gradient as the resolution is further lowered. The true metallicity gradient of the N2S2 \citep{dopita16} metallicity diagnostic is positive and relatively flat at 0.025 dex/Re. The DIG induces a strong negative metallicity gradient of up to -0.21 dex/Re, a deviation of 0.24 dex/Re from the \htwo region metallicity gradient.
\par
Like the N2S2 metallicity gradient, the N2H$\alpha$ \citep{pettini04} metallicity gradient is positive and relatively flat at 0.034 dex/Re. The [NII]/H$\alpha$ emission line ratio is significantly enhanced in the DIG and exceeds the upper limit of the metallicity calibration. Due to the high metallicity and flat gradient of the \htwo regions, any spaxels which are significantly affected by DIG will exceed the upper limit and be removed from the metallicity gradient calculation. For M83, the N2H$\alpha$ metallicity gradient is unaffected by the DIG because of its proximity to the upper limit of the \citet{pettini04} calibration.
\par
As the spatial resolution is decreased from the native 41 pc to 1005 pc, the emission line ratios of the larger region are a luminosity weighted average of the smaller regions within it. To help untangle the evolution of the metallicity gradients with resolution, we present the \htwo dominance profile of M83. The \htwo dominance profile shows the percentage of H$\alpha$ luminosity contributed by \htwo regions as a function of radius. The \htwo dominance profile clearly shows the transition from a numerical weighted average at 41 pc to a luminosity weighted average at 1005 pc. 
\par
Luminosity weighted averaging affects the emission line ratios rather than the metallicities directly. We show the emission line ratio profiles of the \htwo and DIG regions as a function of radius and resolution for each emission line ratio. For all resolution scales, the emission line ratio radial profiles of \htwo regions remains relatively unchanged. However there appears to be significant changes to the DIG emission line ratio profiles for ([OII]+[OIII])/H$\beta$ and [NII]/[OII]. The change in the emission line ratio profiles indicate the significance of low surface brightness regions. As resolution is decreased, more low surface brightness emission is included in the spectrum, leading to significant changes in the emission line ratio profiles of DIG regions. The origin of the low surface brightness DIG emission is still uncertain, but is likely due to a combination of photon leakage from \htwo regions as well as ionization from low-mass evolved stars. The different ionizing mechanisms of the DIG cause the emission line ratios to vary from what we would expect from \htwo regions.
\par
We find that the ([OIII]/H$\beta$)/([NII]/H$\alpha$) emission line ratio radial profile is similar for both high surface brightness DIG and low surface brightness DIG, meaning that decreasing the resolution to kiloparsec resolution scales does not change the overall DIG emission line ratio profile. ([OII]+[OIII])/H$\beta$ and [NII]/[OII] have different emission line ratio radial profiles for the high and low surface brightness DIG, leading to changes in the overall DIG emission line ratio profiles as we decrease the resolution scales. Overall this means that the O3N2 emission line diagnostic will provide the closest metallicity gradient to the pure \htwo region metallicity gradient at a resolution of 1005 pc, the resolution scales of large galaxy surveys. 
\par
Finally we discuss the implications this study has for low resolution observations where the DIG is difficult to remove due to the mixing of \htwo and DIG regions. Based on the DIG characteristics of M83, the emission line radial profiles of DIG tend to possess an opposite gradient to the \htwo regions to varying degrees at a resolution of 1005 pc, the resolution scales of large galaxy surveys.
\par
Applying the [SII]/H$\alpha$ classification scheme at any resolution for all metallicity diagnostics used in this study will yield a metallicity gradient closer to the metallicity gradient of pure \htwo regions, but is unable to fully produce it. The DIG contamination associated with coarser resolution elements is unavoidable, but we are able to remove spaxels which are almost completely comprised of DIG by applying the [SII]/H$\alpha$ cut. 
\par
It is not yet possible to correct for the DIG in large galaxy surveys or high redshift observations without analysing the DIG fractions and DIG emission line properties of multiple galaxies. We therefore recommend that when measuring metallicity gradients at low spatial resolutions, that a [SII]/H$\alpha$ cut is used at resolution scales between 150 - 1005 pc to remove spaxels which are almost completely comprised of DIG. When measuring metallicity gradients at low resolution scales, it will inevitably be affected by DIG and should be used with extreme caution. 
\section*{Acknowledgements}

This paper includes data obtained with the du Pont Telescope at the Las Campanas Observatory, Chile as part of the TYPHOON Program, which has been obtaining optical data cubes for the largest angular-sized galaxies in the southern hemisphere. We thank past and present Directors of The Observatories and the numerous time assignment committees for their generous and unfailing support of this long-term program.
\\
Parts of this research were conducted by the Australian Research Council Centre of Excellence for All Sky Astrophysics in 3 Dimensions (ASTRO 3D), through project number CE170100013.
\\
BG gratefully acknowledges the support of the Australian Research Council as the recipient of a Future Fellowship (FT140101202).
\\
This research has made use of the NASA/IPAC Extragalactic Database (NED), which is operated by the Jet Propulsion Laboratory, California Institute of Technology, under contract with the National Aeronautics and Space Administration.

\bibliographystyle{mnras}
\bibliography{References}

\appendix

\onecolumn
\section{Metallicity Gradient Fits}
\begin{figure}
\centering
\begin{tabular}{|lccccc|}
\multicolumn{6}{c}{[SII]/H$\alpha$ Cut} \\
\hline
Diagnostic & 41pc & 150pc & 330pc & 502pc & 1005pc \\
\hline
\textbf{R23} & & & & & \\
\htwo&-0.072$\pm$0.005&-0.059$\pm$0.004&-0.052$\pm$0.006&-0.049$\pm$0.006&-0.046$\pm$0.007\\
\htwo Low Res&-&-0.050$\pm$0.005&-0.038$\pm$0.008&-0.031$\pm$0.010&-0.002$\pm$0.008\\
Full&-0.072$\pm$0.005&-0.040$\pm$0.005&-0.018$\pm$0.008&-0.020$\pm$0.010&0.002$\pm$0.015\\
DIG&-0.005$\pm$0.010&0.002$\pm$0.007&0.040$\pm$0.009&0.037$\pm$0.012&0.081$\pm$0.019\\
\hline
\textbf{N2O2} & & & & &\\
\htwo&-0.114$\pm$0.006&-0.105$\pm$0.005&-0.097$\pm$0.006&-0.096$\pm$0.007&-0.087$\pm$0.008\\
\htwo Low Res&-&-0.089$\pm$0.006&-0.074$\pm$0.008&-0.062$\pm$0.011&-0.021$\pm$0.014\\
Full&-0.080$\pm$0.005&-0.064$\pm$0.005&-0.044$\pm$0.007&-0.033$\pm$0.010&-0.024$\pm$0.013\\
DIG&-0.019$\pm$0.007&-0.029$\pm$0.005&-0.001$\pm$0.009&0.019$\pm$0.010&0.036$\pm$0.017\\
\hline
\textbf{O3N2} & & & & &\\
\htwo&-0.068$\pm$0.003&-0.058$\pm$0.006&-0.056$\pm$0.007&-0.050$\pm$0.009&-0.063$\pm$0.012\\
\htwo Low Res&-&-0.026$\pm$0.007&-0.016$\pm$0.010&-0.015$\pm$0.012&-0.040$\pm$0.010\\
Full&-0.034$\pm$0.003&-0.001$\pm$0.004&-0.008$\pm$0.006&-0.026$\pm$0.009&-0.043$\pm$0.010\\
DIG&0.002$\pm$0.003&0.029$\pm$0.004&0.024$\pm$0.006&0.015$\pm$0.008&0.009$\pm$0.012\\
\hline
\textbf{N2H$\alpha$} & & & & &\\
\htwo&0.034$\pm$0.002&0.024$\pm$0.004&0.025$\pm$0.006&0.026$\pm$0.007&0.027$\pm$0.009\\
\htwo Low Res&-&0.029$\pm$0.007&0.024$\pm$0.013&0.005$\pm$0.016&0.024$\pm$0.033\\
Full&0.019$\pm$0.001&0.015$\pm$0.004&0.026$\pm$0.007&0.012$\pm$0.009&-0.004$\pm$0.014\\
DIG&-0.015$\pm$0.002&-0.011$\pm$0.004&0.003$\pm$0.007&-0.001$\pm$0.010&0.012$\pm$0.016\\
\hline
\textbf{N2S2} & & & & &\\
\htwo&0.025$\pm$0.003&0.024$\pm$0.005&0.026$\pm$0.009&0.030$\pm$0.010&0.036$\pm$0.013\\
\htwo Low Res&-&-0.002$\pm$0.009&-0.009$\pm$0.013&-0.027$\pm$0.016&-0.011$\pm$0.037\\
Full&-0.090$\pm$0.001&-0.150$\pm$0.003&-0.175$\pm$0.006&-0.210$\pm$0.009&-0.180$\pm$0.014\\
DIG&-0.095$\pm$0.001&-0.151$\pm$0.003&-0.176$\pm$0.005&-0.182$\pm$0.006&-0.181$\pm$0.011\\
\hline
\end{tabular}
\caption{Metallicity gradient fits for Figure \ref{metgradgrid_siicut}}
\end{figure}
\begin{figure}
\centering
\begin{tabular}{|lccccc|}
\multicolumn{6}{c}{\sc{HIIphot}} \\
\hline
Diagnostic & 41pc & 150pc & 330pc & 502pc & 1005pc \\
\hline
\textbf{R23} & & & & & \\
\htwo&-0.064$\pm$0.005&-0.058$\pm$0.006&-0.048$\pm$0.008&-0.052$\pm$0.008&-0.042$\pm$0.009\\
\htwo Low Res&-&-&-&-&-\\
Full&-0.072$\pm$0.005&-0.040$\pm$0.005&-0.018$\pm$0.008&-0.020$\pm$0.010&0.002$\pm$0.015\\
DIG&-0.071$\pm$0.009&-0.016$\pm$0.007&0.012$\pm$0.009&0.008$\pm$0.011&0.052$\pm$0.018\\
\hline
\textbf{N2O2} & & & & &\\
\htwo&-0.075$\pm$0.006&-0.104$\pm$0.006&-0.095$\pm$0.008&-0.118$\pm$0.007&-0.079$\pm$0.010\\
\htwo Low Res&-&-&-&-&-\\
Full&-0.080$\pm$0.005&-0.064$\pm$0.005&-0.044$\pm$0.007&-0.033$\pm$0.010&-0.024$\pm$0.013\\
DIG&-0.067$\pm$0.008&-0.041$\pm$0.006&-0.014$\pm$0.008&-0.005$\pm$0.010&0.040$\pm$0.016\\
\hline
\textbf{O3N2} & & & & &\\
\htwo&-0.034$\pm$0.004&-0.035$\pm$0.007&-0.046$\pm$0.010&-0.042$\pm$0.012&-0.054$\pm$0.014\\
\htwo Low Res&-&-&-&-&-\\
Full&-0.034$\pm$0.003&-0.001$\pm$0.004&-0.008$\pm$0.006&-0.026$\pm$0.009&-0.043$\pm$0.010\\
DIG&0.003$\pm$0.003&0.018$\pm$0.004&0.020$\pm$0.006&0.007$\pm$0.008&0.018$\pm$0.011\\
\hline
\textbf{N2H$\alpha$} & & & & &\\
\htwo&0.025$\pm$0.004&0.009$\pm$0.007&0.040$\pm$0.010&0.040$\pm$0.011&0.046$\pm$0.013\\
\htwo Low Res&-&-&-&-&-\\
Full&0.019$\pm$0.001&0.015$\pm$0.004&0.026$\pm$0.007&0.012$\pm$0.009&-0.004$\pm$0.014\\
DIG&0.002$\pm$0.002&0.005$\pm$0.004&0.034$\pm$0.008&0.021$\pm$0.010&-0.015$\pm$0.013\\
\hline
\textbf{N2S2} & & & & &\\
\htwo&-0.020$\pm$0.004&-0.027$\pm$0.009&-0.008$\pm$0.014&0.010$\pm$0.018&0.019$\pm$0.021\\
\htwo Low Res&-&-&-&-&-\\
Full&-0.090$\pm$0.001&-0.150$\pm$0.003&-0.175$\pm$0.006&-0.210$\pm$0.009&-0.180$\pm$0.014\\
DIG&-0.084$\pm$0.002&-0.144$\pm$0.003&-0.158$\pm$0.005&-0.172$\pm$0.007&-0.171$\pm$0.012\\
\hline
\end{tabular}
\caption{Metallicity gradient fits for Figure \ref{metgradgrid_hiiphot}}
\end{figure}

\begin{figure}
\centering
\begin{tabular}{|lccccc|}
\multicolumn{6}{c}{H$\alpha$ Cut} \\
\hline
Diagnostic & 41pc & 150pc & 330pc & 502pc & 1005pc \\
\hline
\textbf{R23} & & & & & \\
\htwo&-0.065$\pm$0.005&-0.056$\pm$0.006&-0.058$\pm$0.007&-0.050$\pm$0.008&-0.044$\pm$0.009\\
\htwo Low Res&-&-0.040$\pm$0.006&-0.040$\pm$0.010&-0.034$\pm$0.013&0.022$\pm$0.027\\
Full&-0.072$\pm$0.005&-0.040$\pm$0.005&-0.018$\pm$0.008&-0.020$\pm$0.010&0.002$\pm$0.015\\
DIG&-0.074$\pm$0.018&0.008$\pm$0.012&0.029$\pm$0.013&0.043$\pm$0.012&0.063$\pm$0.021\\
\hline
\textbf{N2O2} & & & & &\\
\htwo&-0.065$\pm$0.006&-0.085$\pm$0.005&-0.076$\pm$0.007&-0.080$\pm$0.007&-0.077$\pm$0.009\\
\htwo Low Res&-&-0.066$\pm$0.007&-0.063$\pm$0.009&-0.033$\pm$0.013&0.007$\pm$0.032\\
Full&-0.080$\pm$0.005&-0.064$\pm$0.005&-0.044$\pm$0.007&-0.033$\pm$0.010&-0.024$\pm$0.013\\
DIG&-0.092$\pm$0.012&-0.057$\pm$0.012&-0.042$\pm$0.012&0.001$\pm$0.013&0.100$\pm$0.022\\
\hline
\textbf{O3N2} & & & & &\\
\htwo&-0.022$\pm$0.003&-0.008$\pm$0.005&-0.007$\pm$0.007&-0.033$\pm$0.009&-0.042$\pm$0.011\\
\htwo Low Res&-&0.014$\pm$0.009&0.007$\pm$0.012&-0.016$\pm$0.019&0.009$\pm$0.036\\
Full&-0.034$\pm$0.003&-0.001$\pm$0.004&-0.008$\pm$0.006&-0.026$\pm$0.009&-0.043$\pm$0.010\\
DIG&0.017$\pm$0.004&0.028$\pm$0.004&0.020$\pm$0.007&0.021$\pm$0.008&0.014$\pm$0.012\\
\hline
\textbf{N2H$\alpha$} & & & & &\\
\htwo&0.023$\pm$0.002&0.006$\pm$0.004&0.001$\pm$0.006&0.008$\pm$0.008&0.038$\pm$0.011\\
\htwo Low Res&-&0.003$\pm$0.010&0.021$\pm$0.016&0.035$\pm$0.023&-0.173$\pm$0.048\\
Full&0.019$\pm$0.001&0.015$\pm$0.004&0.026$\pm$0.007&0.012$\pm$0.009&-0.004$\pm$0.014\\
DIG&-0.022$\pm$0.002&-0.023$\pm$0.004&-0.006$\pm$0.008&-0.016$\pm$0.010&-0.018$\pm$0.017\\
\hline
\textbf{N2S2} & & & & &\\
\htwo&-0.022$\pm$0.003&-0.025$\pm$0.005&0.008$\pm$0.009&0.004$\pm$0.011&0.013$\pm$0.014\\
\htwo Low Res&-&-0.032$\pm$0.009&-0.032$\pm$0.013&-0.062$\pm$0.018&-0.076$\pm$0.040\\
Full&-0.090$\pm$0.001&-0.150$\pm$0.003&-0.175$\pm$0.006&-0.210$\pm$0.009&-0.180$\pm$0.014\\
DIG&-0.082$\pm$0.002&-0.143$\pm$0.003&-0.167$\pm$0.005&-0.159$\pm$0.006&-0.180$\pm$0.011\\
\hline
\end{tabular}
\caption{Metallicity gradient fits for Figure \ref{metgradgrid_hacut}}
\end{figure}
\begin{landscape}
	\begin{figure}
		\centering
		\includegraphics[width=1.3\textheight]{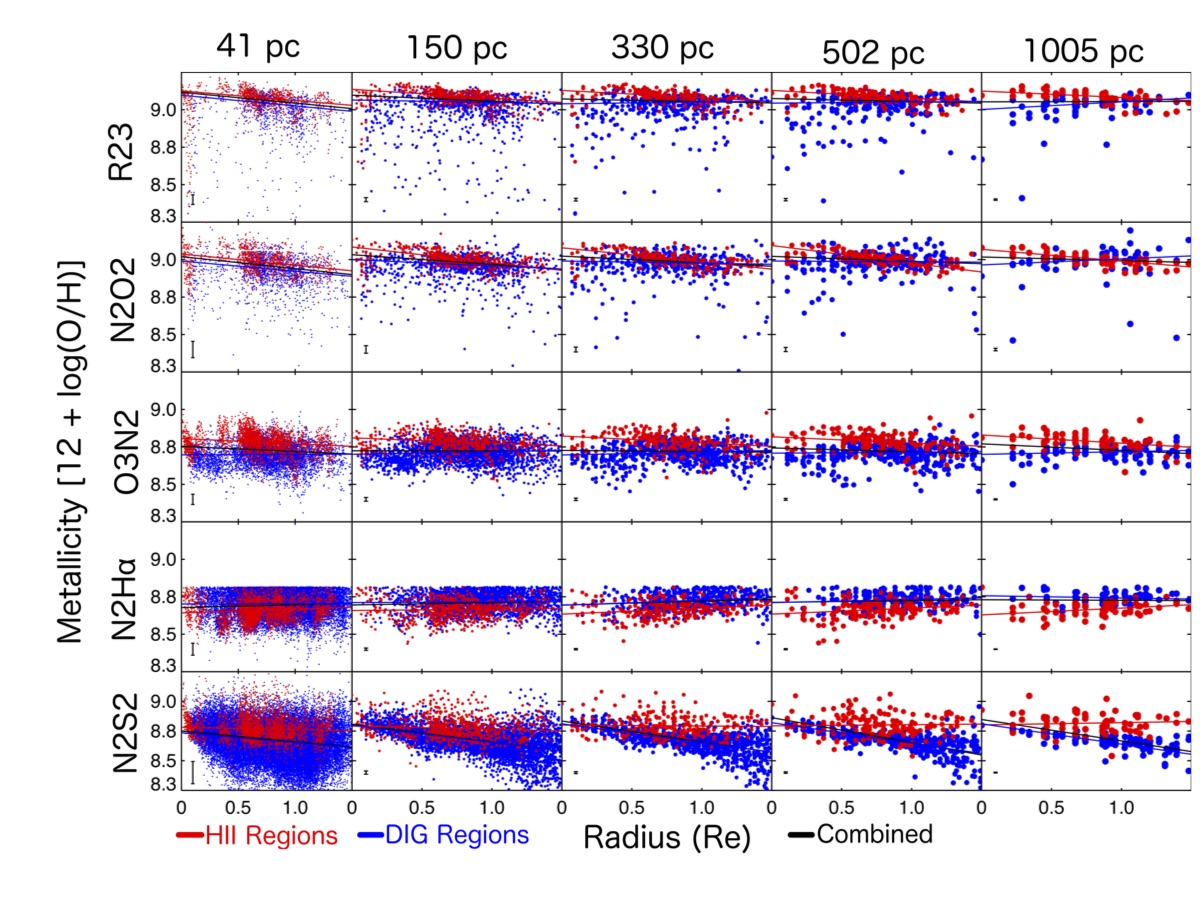}\label{fig:a}
		\caption{
Same as Figure \ref{metgradgrid_siicut} using the \htwo regions defined by {\sc HIIphot}.
		}
		\label{metgradgrid_hiiphot}
	\end{figure}
\end{landscape}

\begin{landscape}
	\begin{figure}
		\centering
		\includegraphics[width=1.3\textheight]{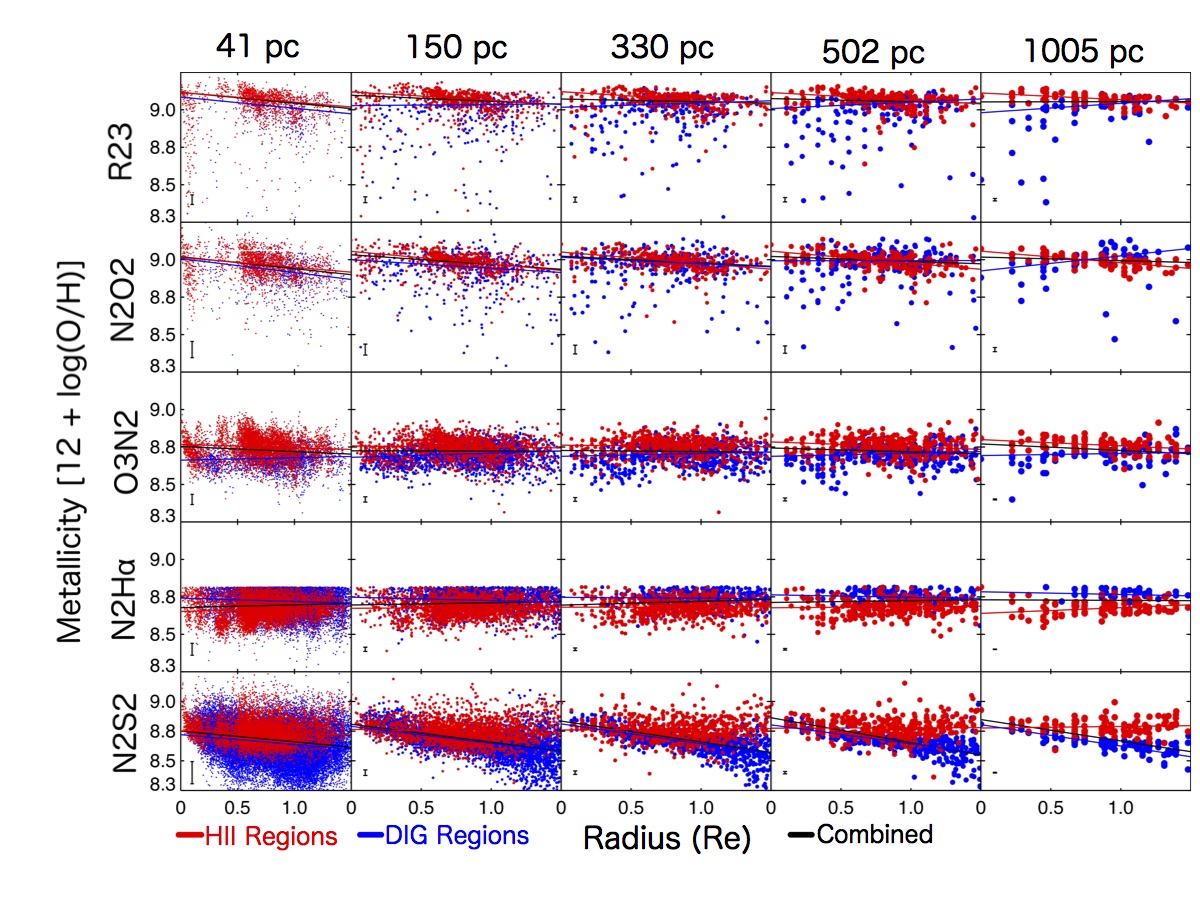}\label{fig:a}
		\caption{
Same as Figure \ref{metgradgrid_siicut} using the \htwo regions defined by the H$\alpha$ surface brightness cut-off.
		}
		\label{metgradgrid_hacut}
	\end{figure}
\end{landscape}
\section{Defining \htwo regions based on Kaplan et al. 2016}
\label{fitvalues}
\begin{figure}
\centering
\includegraphics[width=0.5\textwidth]{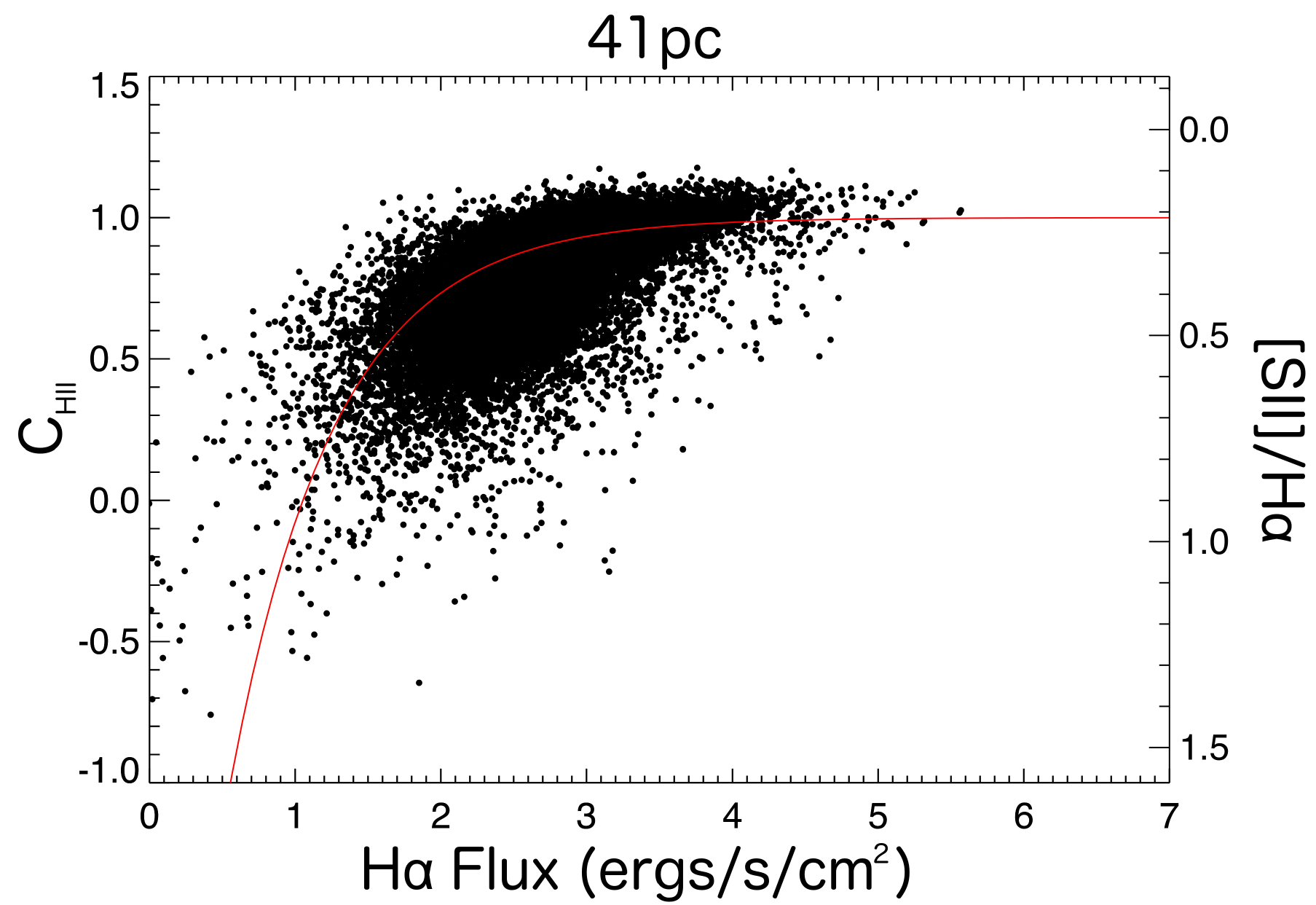}
\caption{The relationship between [SII]/H$\alpha$ and the extinction corrected H$\alpha$ surface brightness. We use the method by \citet{kaplan16} to determine the amount of flux originating from \htwo regions and DIG regions at the native resolution.}
\label{41pcs2ha}
\end{figure}
\begin{figure}
\centering
\includegraphics[width=0.5\textwidth]{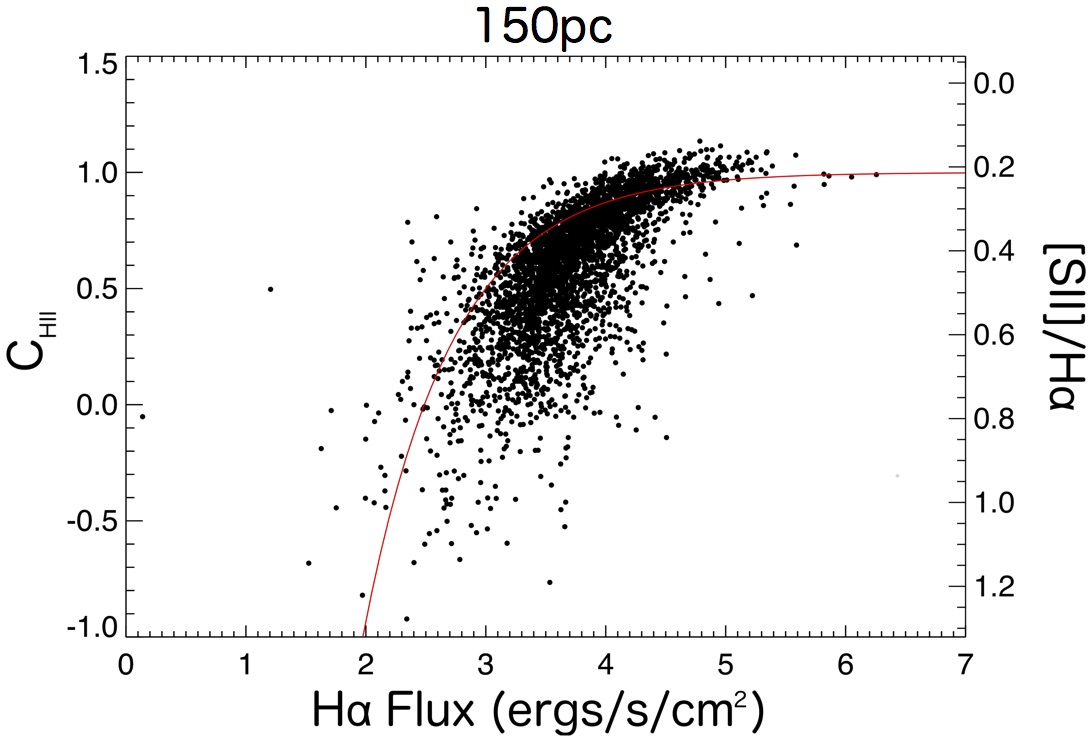}
\caption{Calculation of the \htwo region fraction at 150 pc}
\label{150pcs2ha}
\end{figure}
\begin{figure}
\centering
\includegraphics[width=0.5\textwidth]{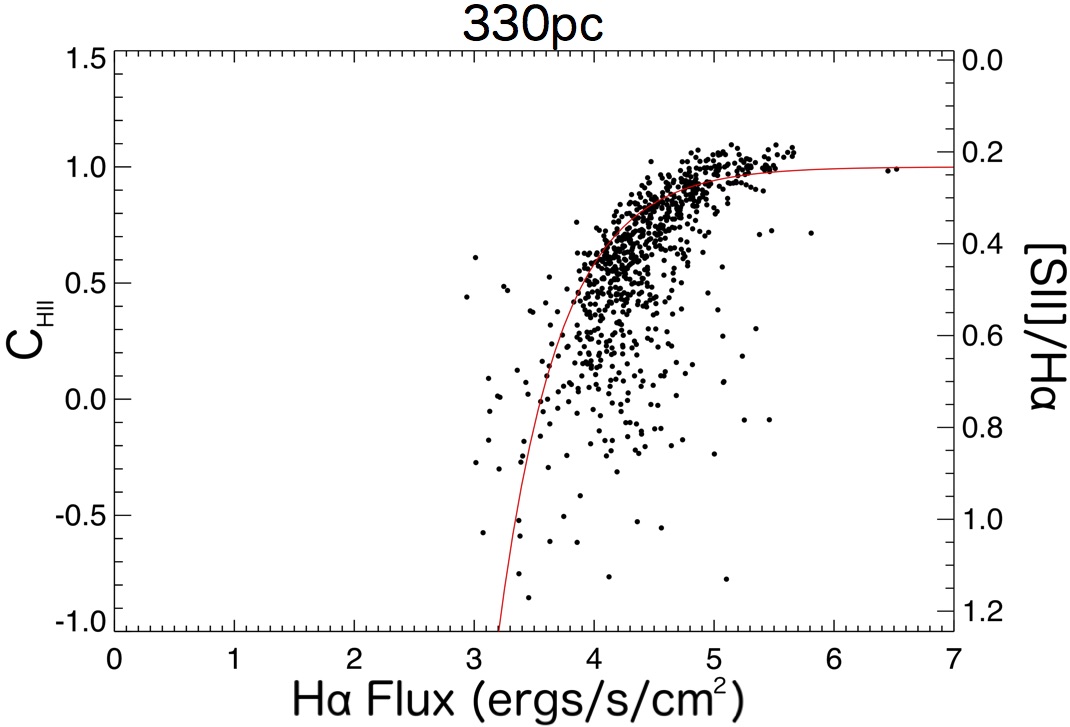}
\caption{Calculation of the \htwo region fraction at 330 pc}
\label{330pcs2ha}
\end{figure}
\begin{figure}
\centering
\includegraphics[width=0.5\textwidth]{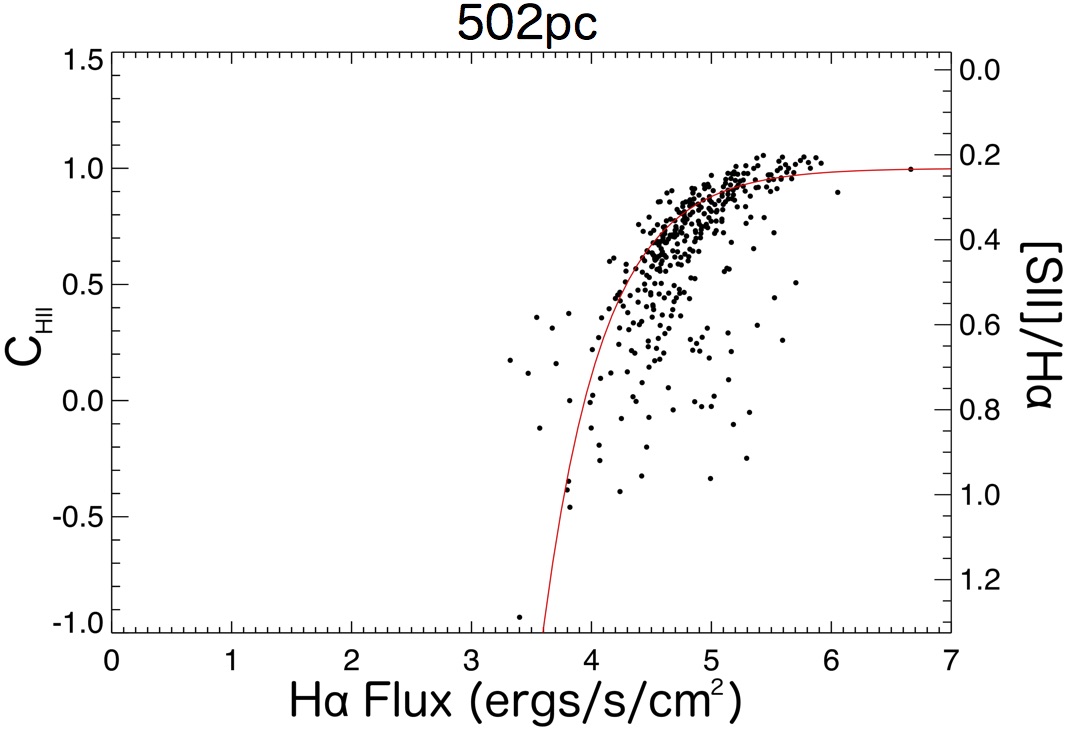}
\caption{Calculation of the \htwo region fraction at 502 pc}
\label{502pcs2ha}
\end{figure}
\begin{figure}
\centering
\includegraphics[width=0.5\textwidth]{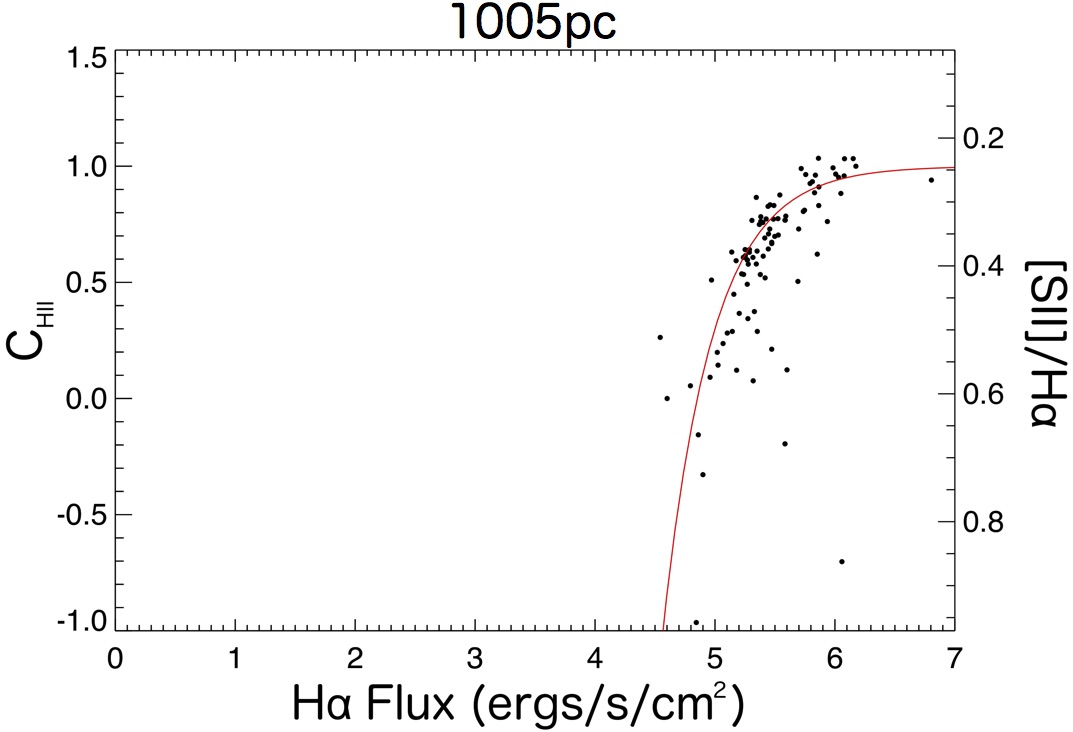}
\caption{Calculation of the \htwo region fraction at 1005 pc}
\label{1005pcs2ha}
\end{figure}
\begin{figure}
\centering
\begin{tabular}{|c|c|c|}
\hline
\multicolumn{3}{|c|}{DIG Fraction Parameters} \\
\hline
Resolution & $\log(f_{0})$ & $\beta$ \\
\hline
41 pc, Figure \ref{41pcs2ha}& 1.05 &0.61 \\
150 pc& 2.49   & 0.59 \\
330 pc &3.56 & 0.85 \\
502 pc   &3.94 & 0.87 \\
1005 pc& 4.85 & 1.05 \\
\hline
\end{tabular}
\caption{Fit parameters for Equation 1 at each resolution}
\end{figure}

\bsp	\label{lastpage}
\end{document}